\documentclass[draft]{agujournal2019}
\usepackage{url} 
\usepackage[inline]{trackchanges} 
\usepackage{soul}
\usepackage{aas_macros}
\justifying

\journalname{Icarus}

\begin{document}

%
%


\title{Latitudinal variation of methane mole fraction above clouds in Neptune's atmosphere from VLT/MUSE-NFM: Limb-darkening reanalysis}

%
%




\authors{Patrick G. J. Irwin\affil{1}, Jack Dobinson\affil{1}, Arjuna James\affil{1}, Daniel Toledo\affil{2}, Nicholas A. Teanby\affil{3}, Leigh N. Fletcher\affil{4}, Glenn S. Orton\affil{5}, Santiago P\'{e}rez-Hoyos\affil{6}}


\affiliation{1}{Department of Physics, University of Oxford, Parks Rd, Oxford, OX1 3PU, UK}
\affiliation{2}{Instituto Nacional de T\'ecnica Aeroespacial (INTA), 28850, Torrej\'on de Ardoz (Madrid), Spain.}
\affiliation{3}{School of Earth Sciences, University of Bristol, Wills Memorial Building, Queens Road, Bristol, BS8 1RJ, UK}
\affiliation{4}{School of Physics \& Astronomy, University of Leicester, University Road, Leicester, LE1 7RH, UK}
\affiliation{5}{Jet Propulsion Laboratory, California Institute of Technology, 4800 Oak Grove Drive, Pasadena, CA 91109, USA}
\affiliation{6}{University of the Basque Country UPV/EHU, 48013 Bilbao, Spain}




\correspondingauthor{Patrick Irwin}{patrick.irwin@physics.ox.ac.uk}




\begin{keypoints}
\item Minnaert limb-darkening analysis improves modelling of Neptune’s reflectivity spectrum in visible/near-IR.
\item General cloud distribution can be modelled with zonally-symmetric H$_2$S cloud and stratospheric haze.
\item Mole fraction of methane at 2--4 bar (above H$_2$S cloud) found to decrease from 4--6\% at the equator to 2--4\% at the south pole.
\item Discrete cloud features can be fitted with an additional methane ice cloud at pressures less than $\sim $ 0.4 bar.

\end{keypoints}

%
%

%
%


\begin{abstract}
We present a reanalysis of visible/near-infrared (480 -- 930 nm) observations of Neptune, made in 2018 with the Multi Unit Spectroscopic Explorer (MUSE) instrument at the Very Large Telescope (VLT) in Narrow Field Adaptive Optics mode, reported by Irwin et al., Icarus, 311, 2019. 
We find that the inferred variation of methane abundance with latitude in our previous analysis, which was based on central meridian observations only, underestimated the retrieval errors when compared with a more complete assessment of Neptune’s limb darkening. In addition, our previous analysis introduced spurious latitudinal variability of both the abundance and its uncertainty, which we reassess here. Our reanalysis of these data incorporates the effects of limb-darkening based upon the Minnaert approximation model, which provides a much stronger constraint on the cloud structure and methane mole fraction, makes better use of the available data and is also more computationally efficient. We find that away from discrete cloud features, the observed reflectivity spectrum from 800 -- 900 nm is very well approximated by a background cloud model that is latitudinally varying, but zonally symmetric, consisting of a H$_2$S cloud layer, based at 3.6 -- 4.7 bar with variable opacity and scale height, and a stratospheric haze. The background cloud model matches the observed limb darkening seen at all wavelengths and latitudes  and we find that the mole fraction of methane at 2--4 bar, above the H$_2$S cloud, but below the methane condensation level, varies from 4--6\% at the equator to 2--4\% at south polar latitudes, consistent with previous analyses, with a equator/pole ratio of $1.9 \pm 0.2$ for our assumed cloud/methane vertical distribution model. The spectra of discrete cloudy regions are fitted, to a very good approximation, by the addition of a single vertically thin methane ice cloud with opacity ranging from 0 -- 0.75 and pressure less than $\sim 0.4$ bar.
\end{abstract}


%
%

%


%
%
%
%

\section{Introduction}

The visible and near-infrared spectrum of Neptune is formed by the reflection of sunlight from the atmosphere, modulated primarily by the absorption of gaseous methane, but also to a lesser extent H$_2$S \cite{irwin18}. Measured spectra can thus be inverted to determine the cloud structure as a function of location and altitude, providing we know the vertical and latitudinal distribution of methane. Although for some years the vertical profiles of methane determined from Voyager 2 radio-occultation observations were used at all latitudes, HST/STIS observations of Uranus recorded in 2002 \cite{kark09} and similar observations of Neptune recorded in 2003 \cite{kark11} both showed that the tropospheric cloud-top (i.e., above the H$_2$S cloud) methane mole fraction  varies significantly with latitude on both planets, later confirmed for Uranus by several follow-up studies \cite{sromovsky11, sromovsky14, sromovsky19}.  These HST/STIS observations used the collision-induced absorption (CIA) bands of H$_2$--H$_2$ and H$_2$--He near 825 nm, which allow variations of CH$_4$ mole fraction to be differentiated from cloud-top pressure variations of the H$_2$S cloud. \citeA{kark09,kark11} found that the methane mole fraction above the main observable H$_2$S cloud tops at 2--4 bar varies from $\sim4$\% at equatorial latitudes to $\sim2$\% polewards of $\sim 40^\circ$ N,S for both planets.

More recently, an analysis of VLT/MUSE Narrow-Field Mode (NFM) observations (770 -- 930 nm) along Neptune's central meridian \cite{irwin19} found a similar latitudinal variation of cloud-top methane mole fraction, with values of 4--5\% reported at equatorial latitudes, reducing to 3--4\% at polar latitudes, but with considerable pixel-to-pixel variation that was not understood at the time. In this study we reanalyse these data using a new limb-darkening approximation model, which makes much better use of all the data from a given latitude, observed at many different zenith angles, and which we find considerably improves our methane mole fraction determinations. We also find that having constrained the smooth latitudinal variation of opacity of the tropospheric cloud and stratospheric haze, we are able to efficiently retrieve the additional opacity of discrete upper tropospheric (0.1 - 0.5 bar) methane clouds seen in our observations.  

\section{MUSE Observations} \label{observations}

As reported by \citeA{irwin19}, commissioning-mode observations of Neptune were made on 19th June 2018 with the Multi Unit Spectroscopic Explorer (MUSE) instrument \cite{bacon10} at ESO's Very Large Telescope (VLT) in Chile, in Narrow-Field Mode (NFM). MUSE is an integral-field spectrograph, which records 300 $\times$ 300 pixel `cubes', where each `spaxel' contains a complete visible/near-infrared spectrum (480 -- 930 nm) with a spectral resolving power of 2000 -- 4000. MUSE's  Narrow-Field Mode has a field of view of 7.5" $\times$ 7.5", giving a spaxel size of 0.025", and uses Adaptive Optics to achieve a spatial resolution less than 0.1".  These commissioning observations are summarised in Table 1 of \citeA{irwin19}. The spatial resolution was estimated to have a full-width-half-maximum of 0.06" at 800 nm. The observed spectra were smoothed to the resolution of the IRTF/SpeX instrument, which has a triangular instrument function with FWHM = 2 nm, sampled at 1 nm, in order to increase the signal-to-noise ratio without losing the essential shape of the observed spectra. This resolution was also more consistent with the spectral resolution of the methane gaseous absorption data used, which are described in section 3.2.

In our previous analysis of these data \cite{irwin19}, spectra recorded from single pixels along the central meridian of one of the longer integration time observations (120s) were fitted with our NEMESIS  retrieval model \cite{irwin08} to determine latitudinal variations of methane and cloud structure. The cloud-top (i.e., immediately above the H$_2$S cloud) methane mole fractions were found to be consistent with HST/STIS observations \cite{kark11}, but were not very well constrained with significant latitudinal variation that we attributed at the time to the random noise from single-pixel retrievals. However, they also did not make full use of the limb-darkening behaviour visible in these IFU observations, although we verified that our cloud parameterization reproduced the observed limb-darkening well at 5 -- $10^\circ$S. 

Since making our initial report on the MUSE-NFM Neptune observations, an analysis of HST/WFC3 observations for Jupiter has been conducted by \citeA{perez20}, which makes much better use of the limb-darkening information content of multi-spectral observations using a  Minnaert limb darkening approximation scheme. We have adapted this technique for use with our Neptune MUSE-NFM observations and find that it greatly improves the quality of our fits and our estimates of the latitudinal variation of cloud-top methane mole fraction at 2--4 bar in Neptune's atmosphere. This reanalysis has also highlighted an erroneous retrieval artefact in our previous work \cite{irwin19} at some locations, which can now be explained.

\section{Reanalysis}\label{analysis}

\subsection{Minnaert Limb-darkening analysis}\label{minnaert}
 
 The dependence of the observed reflectivity from a location on a planet on the incidence and emission angles can be well approximated using an empirical law first introduced by \citeA{minnaert41}. For an observation at a particular wavelength, the observed reflectivity\footnote{$I/F$ is $\pi R/F$, where $R$ is the reflected radiance (W cm$^{-2}$ sr$^{-1}$ $\mu$m$^{-1}$) and $F$ is the incident solar irradiance at the planet (W cm$^{-2}$ $\mu$m$^{-1}$). }  $I/F$  can be approximated as:
 
\begin{linenomath*}
\begin{equation}\label{eq:minnaert}
\frac{I}{F} = \left(\frac{I}{F}\right)_{0}  \mu_{0}^{k} \mu^{k-1}
\end{equation}
\end{linenomath*}

where $(I/F)_0$ is the nadir-viewing reflectivity, $k$ is the limb-darkening parameter, and $\mu$ and $\mu_0$ are, respectively, the cosines of the emission and solar incidence angles. With this model a value of $k > 0.5$ indicates limb-darkening, while $k < 0.5$ indicates limb-brightening.    Taking logarithms,  Eq. \ref{eq:minnaert} can be re-expressed as:

\begin{linenomath*}
\begin{equation}\label{eq:minnaert_logs}
\ln\left(\mu\frac{I}{F}\right) =\ln\left(\frac{I}{F}\right)_{0}+k \ln(\mu\mu_0)
\end{equation}
\end{linenomath*}

and we can see that it is possible to fit the Minnaert parameters $(I/F)_0$ and $k$ if we perform a least-squares fit on a set of measurements of $\ln(\mu I/F)$ as a function of  $\ln(\mu\mu_0)$. 

\begin{figure} 
\includegraphics[width=\textwidth]{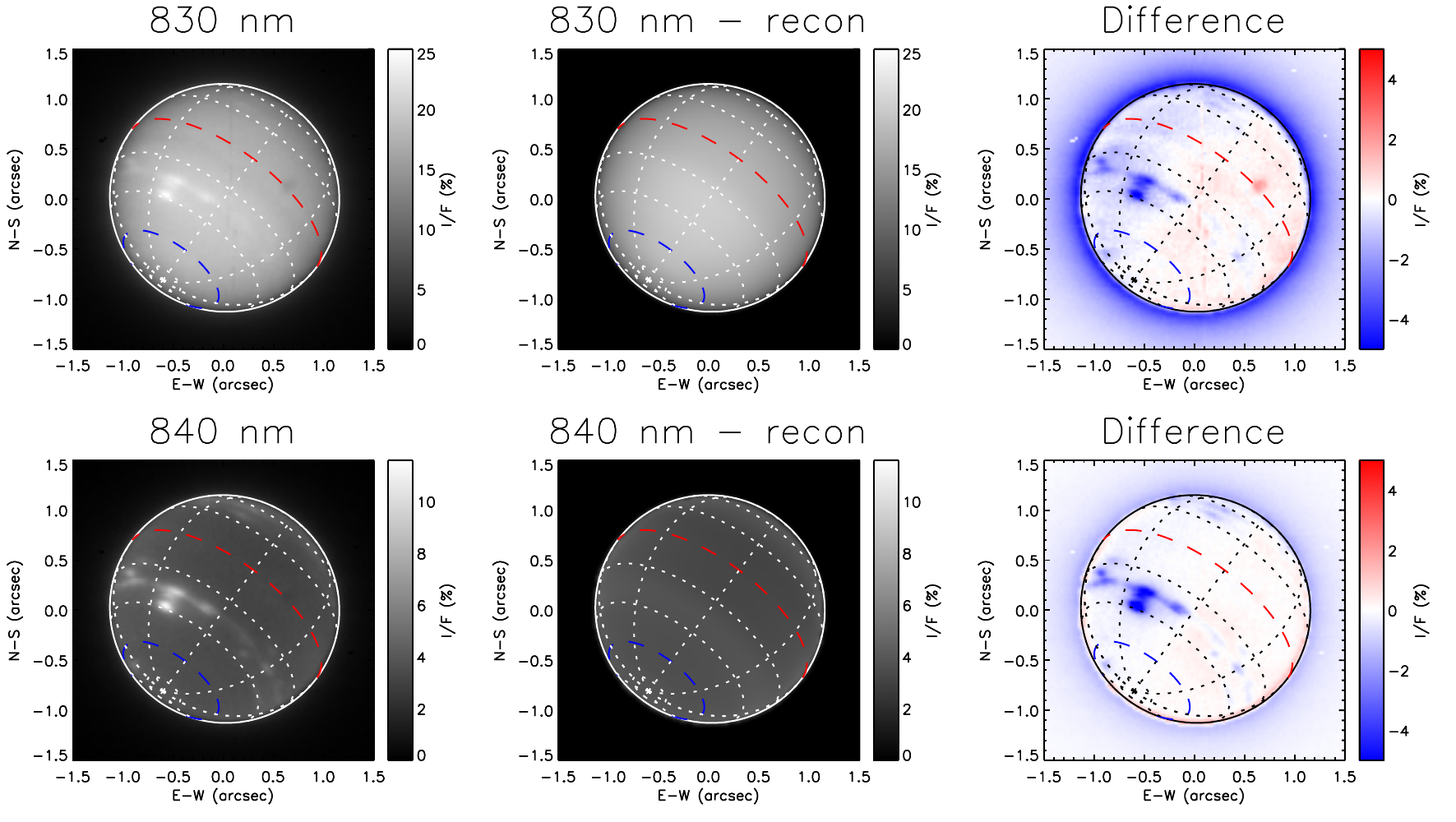}
\caption{Appearance of Neptune on June 19th 2018 at 09:43:21UT (Observation `3' of \citeA{irwin19}) at 830 nm (top row) and 840 nm (bottom row). The left-hand column shows the observed images and the middle column shows the images reconstructed following a Minnaert limb-darkening analysis. Here, for each location with latitude $\phi$ and cos(zenith) angles $\mu_0$ and $\mu$ the reflectivity was calculated as $ I/F = R(\phi)  \mu_{0}^{k(\phi)} \mu^{k(\phi)-1}$ where $R(\phi)$ and $k(\phi)$ are the interpolated values of Minnaert parameters $\left(I/F\right)_{0}$ and $k$ at that latitude. The right-hand column shows the difference (reconstructed - observed). The equator and $60^\circ$S latitude circles are indicated by the red-dashed and blue-dashed lines, respectively. The dark spot seen at continuum wavelengths near the equator is an artefact of the reduction pipeline of unknown origin. We found it to have negligible effect on our analysis.} \label{fig:neptune_recon}
\end{figure}

\begin{figure} 
\includegraphics[width=0.5\textwidth]{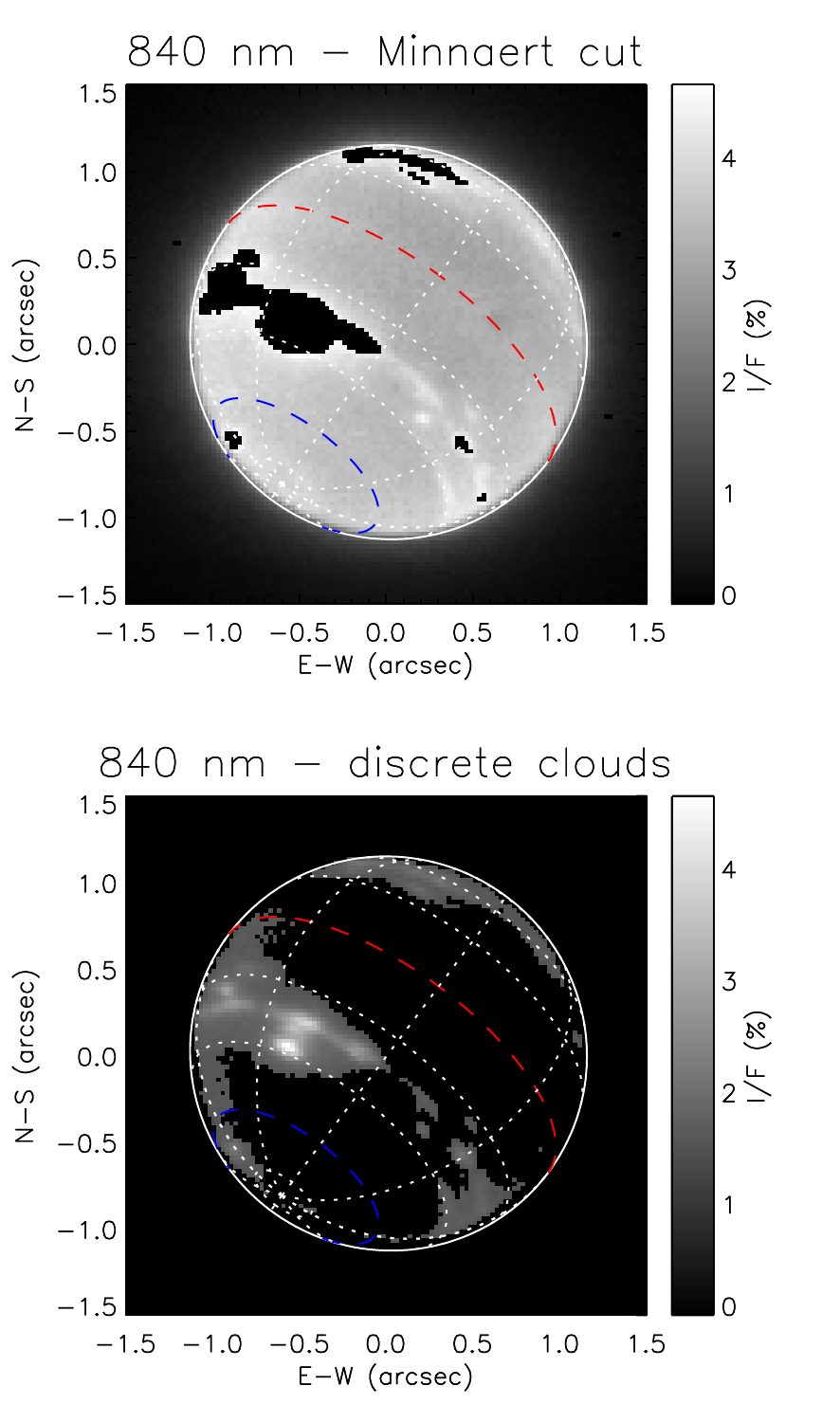}
\caption{Appearance of Neptune on June 19th 2008 at 09:43:21UT (Observation `3' of \citeA{irwin19}) at 840 nm, showing regions masked to fit the Minnaert curves (top) and the areas selected for later additional discrete cloud fitting (bottom).} \label{fig:neptune_cut}
\end{figure}

\begin{figure} 
\includegraphics[width=\textwidth]{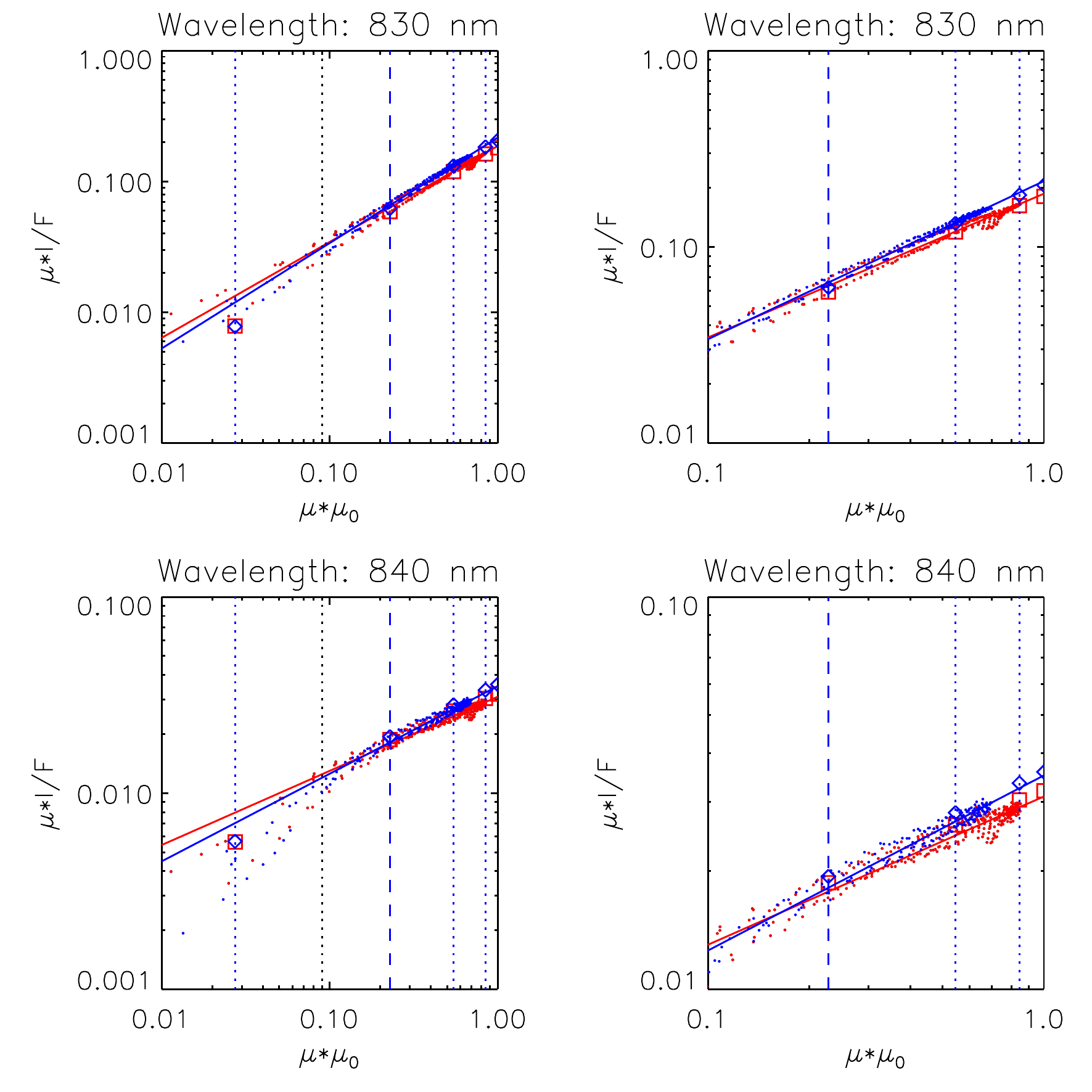}
\caption{Minnaert analysis of observed limb-darkening curves at 830 nm (top row) and 840 nm (bottom row). The left-hand column shows all observations, while the right-hand column is limited to points with $\mu\mu_0 > 0.1$. The red points are the measurements at the equator, while those coloured blue are at $60^\circ$S. The red and blue solid lines are the lines fitted with the Minnaert analysis for points with $\mu\mu_0 \ge 0.09$. The vertical blue dotted lines indicate the values of $\mu\mu_0$ corresponding to our 5-point quadrature ordinates (with the vertical blue dashed line indicating the smallest value of $\mu\mu_0$ used in our retrieval analysis). The blue diamonds and red squares are simulated reflectivities from our best-fit cloud structure and methane mole fractions at these latitudes, reported in section 3.3, showing excellent consistency between the Minnaert approximation and the reflectivities simulated with our matrix operator multiple-scattering model for $\mu\mu_0 > 0.1$.} \label{fig:limbcurve}
\end{figure}

We analysed the same `cube' of Neptune as was studied by \citeA{irwin19}, namely Observation `3', recorded by VLT/MUSE at 09:43:21(UT) on 19th June 2018. We analysed the spectra in this cube in the wavelength range 800 -- 900 nm and Fig.~\ref{fig:neptune_recon} shows the observed appearance of the planet at 830 and 840 nm, which are wavelengths of weak and strong methane absorption, respectively. The limb-darkening behaviour of the observed spectra were analysed in latitude bands of width $10^\circ$, spaced every $5^\circ$ to achieve Nyquist sampling. For each latitude band, the observed reflectivities were used to construct plots of $\ln(\mu I/F)$ against $\ln(\mu\mu_0)$ and straight lines fitted to deduce $(I/F)_0$ and $k$ for each wavelength. Locations on the disc where there were bright clouds were masked out (Fig. \ref{fig:neptune_cut}) and examples of the fits at 830 and 840 nm  for latitude bands centred on the equator and $60^\circ$S are shown in Fig.~\ref{fig:limbcurve}. Here it can be seen that the Minnaert empirical law provides a very accurate approximation of the observed dependence of reflectivity with viewing zenith angles. Although all the measurements are plotted, only those measurements with $\mu\mu_0 \geq 0.09$ (i.e., $\mu$, $\mu_0 > \sim 0.3$) were used to fit $(I/F)_0$ and $k$ to make sure that the fitting procedure was not overly affected by points measured near the disc edge and thus potentially more `diluted' with space. Also plotted in Fig. \ref{fig:limbcurve} are the reflectivities calculated with our radiative transfer and retrieval model from our best-fit retrieved cloud and methane mole fractions at these latitudes, reported in section 3.3. It can be seen that there is very good agreement between the reflectivities calculated with our multiple-scattering matrix operator model and the Minnaert limb-darkening approximation to the observations for zenith angles less than $\sim 70^\circ$. 

\begin{figure} 
\includegraphics[width=\textwidth]{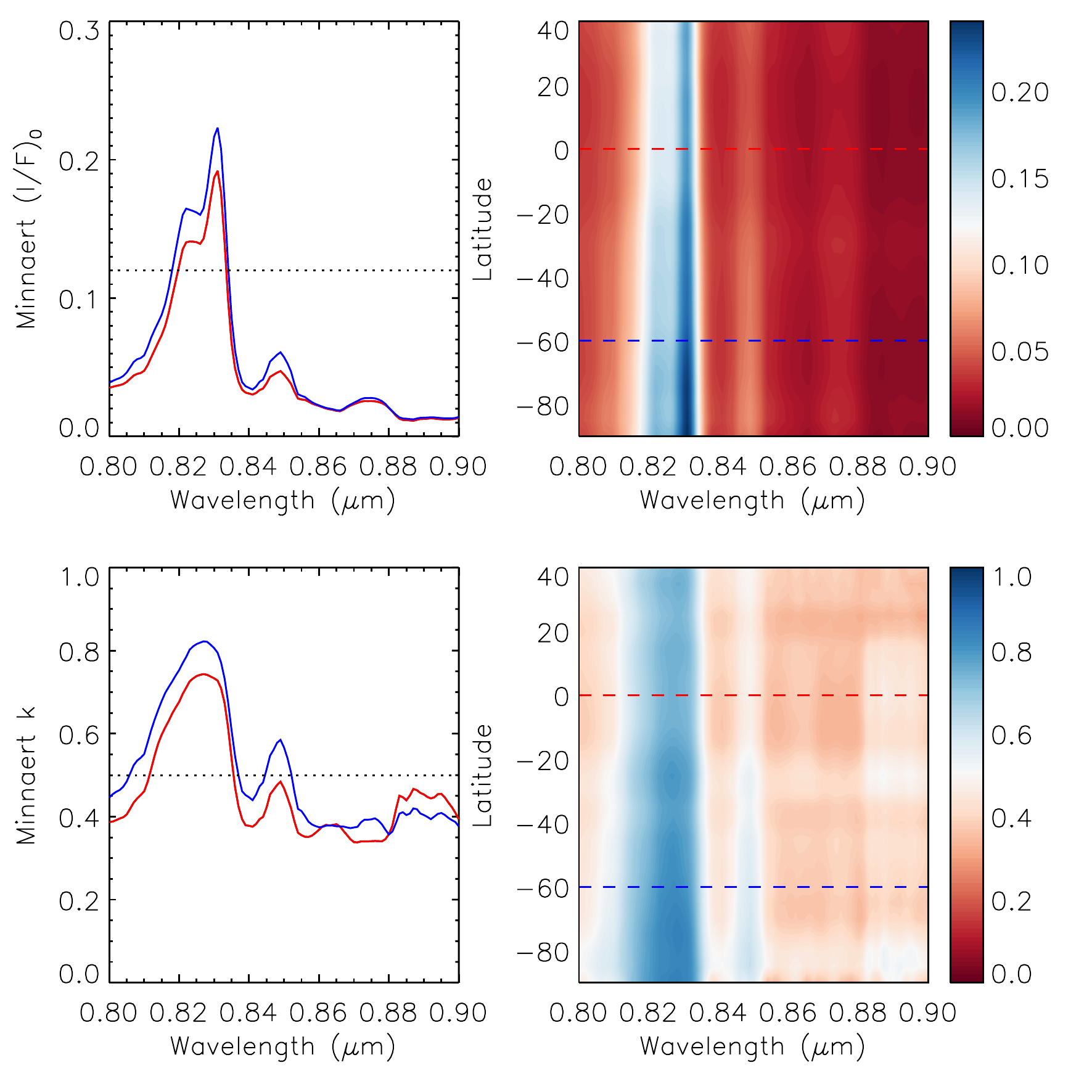}
\caption{Minnaert analysis of observed limb-darkening curves from 800 to 900 nm and at all latitudes visible on Neptune's disc. The left-hand column shows the fitted values of $(I/F)_0$ and $k$ at the equator (red) and $60^\circ$S (blue), while the right-hand column shows a contour plot of these fitted parameters at all latitudes. The horizontal lines in the left hand column indicates the value of the white colour in the contour plots in the right hand column, and for the $k$ contour plot indicates the transition from limb darkening $(k > 0.5)$ to limb-brightening $(k < 0.5)$. The horizontal lines in the right hand column indicate the equator (red) and $60^\circ$S (blue), respectively.} \label{fig:spec_contour}
\end{figure}

Extending this analysis to all wavelengths under consideration, Fig.~\ref{fig:spec_contour} shows a contour plot of the fitted values of $(I/F)_0$ and $k$ for all wavelengths and latitude bands. It can be seen that at wavelengths near 830 nm, the fitted $k$ values are greater than 0.5, indicating limb darkening, while at longer wavelengths, values of $k$ less than 0.5 are fitted, indicating limb brightening. It can also just be seen in Fig.~\ref{fig:spec_contour} that the width of the reflectance peak of $(I/F)_0$ is noticeably wider at latitudes southwards of 20 -- $40^\circ$S, a trend that is also just discernible in the fitted $k$ values near the reflectance peak.

Having fitted values of $(I/F)_0$ and $k$ for all wavelengths and latitude bands, it is then possible to reconstruct the apparent image of the planet at any observation geometry. Using the measured observation values of $\mu$ and $\mu_0$ we reconstructed the images of Neptune at 830 and 840 nm, where for each location with latitude $\phi$ and cos(zenith) angles $\mu_0$ and $\mu$ the reflectivity is calculated as $I/F = R(\phi) \mu_{0}^{k(\phi)} \mu^{k(\phi)-1}$, where $R(\phi)$ and $k(\phi)$ are the interpolated values of Minnaert parameters $\left(I/F\right)_{0}$ and $k$ at that latitude. We compare these reconstructed images with the observed images in Fig.~\ref{fig:neptune_recon}, which also shows the differences between the observed and reconstructed images. As can be seen the observed general dependence of reflectivity with latitude and position on disc is well reproduced compared with the original MUSE observations and the differences are very small, except at: 1) locations of known artefacts in the reduced data; 2) locations of the small discrete clouds, which were masked out when fitting the zonally-averaged Minnaert limb-darkening curves; and 3) off-disc, where the observed images are not corrected for the instrument point-spread function (PSF). We will return to these discrete clouds in section 3.5.  

\subsection{Retrieval model}\label{retrieval_model}

Having applied the Minnaert model to the observations we then used the fitted $(I/F)_0$ and $k$ parameters to reconstruct synthetic spectra of Neptune for all visible latitude bands and fitted these as synthetic `observations' using our radiative transfer and retrieval model, NEMESIS \cite{irwin08}.  There are two main advantages in doing this: 1) the spectra reconstructed using the fitted Minnaert parameters have smaller random error values as they have been reconstructed from values fitted to a combination of all the points in a latitude band; and 2) we can reconstruct the apparent spectrum of Neptune at any set of angles that is convenient for modelling, which can greatly reduce computation time. In our previous approach \cite{irwin19}, where we did not assume to know the zenith-angle dependence, we tried to fit simultaneously to the observations at several different zenith angles near the equator. For modelling near-infrared reflectivity observations, NEMESIS employs a plane-parallel matrix operator multiple-scattering model \cite{plass73}. In this model, integration over zenith angle is done with a Gauss-Lobatto quadrature scheme, while the azimuth integration is done with Fourier decomposition. For most calculations, not too near to the disc edge, we have found that five zenith angles are usually sufficient and the reflectivity at a particular zenith angle is linearly interpolated between calculations done at the two closest zenith angles. Although this provides a general purpose functionality, this approach has some drawbacks: 1) it requires two sets of calculations at two different solar zenith angles for each location; 2) the linear interpolation can lead to interpolation errors at larger zenith angles; and 3) for points near the disc edge, the number of Fourier components needed to fully resolve the azimuth dependence increases, which can greatly increase computation time. By reconstructing spectra using the fitted Minnaert $(I/F)_0$ and $k$ parameters we can simulate spectra measured as if they exactly coincided with the angles in our quadrature scheme, thus avoiding interpolation error. In addition, if we assume the Minnaert approximation to be true, which has a linear dependence in logarithmic space, we only need to fit spectra calculated at two different angles, not several, and can test how well the linear approximation applies in post-processing. Hence, in the retrievals presented here we reconstructed two spectra for each latitude band with viewing zenith angle $\theta_V$, solar zenith angle $\theta_S$ and azimuth angle $\phi$ values of ($0^\circ$, $0^\circ$, $180^\circ$) and ($61.45^\circ$, $61.45^\circ$, $180^\circ$), respectively. Here, $\phi = 180^\circ$ indicates back-scattering, while $\theta = 61.45^\circ$ is the second zenith angle in our five-point Gauss-Lobatto scheme, summarised in Table~\ref{tab:quadrature}, and $\theta = 0^\circ$ is the fifth. This second zenith angle is sufficiently high to probe limb-darkening or limb-brightening, but is not too high that we need an excessive number of Fourier components in the azimuth angle decomposition to properly model it, which would make the computation excessively slow. For each latitude band the two synthetic observations at $0^\circ$ and $61.45^\circ$ zenith angle were then fitted simultaneously to determine the vertical cloud structure and tropospheric methane mole fraction.

Errors in the fitted $(I/F)_0$ and $k$ parameters were propagated into the errors on these reconstructed spectra at $0^\circ$ and $61.45^\circ$ zenith angle as normal, and were seen to increase towards the poles where the curves were less well sampled. However, even then, because the synthetic spectra are derived from linear fits to a large number of data points the random error is very small and we found that we were unable to fit the synthetic observations to within these error. We attribute this to `forward-modelling' systematic errors due to deficiencies in the methane absorption coefficients of \citeA{kark11} and also in our chosen cloud and methane parameterization schemes, described below. In order to achieve final $\chi^2/n$ fits of $\sim 1$ at all latitudes in our retrievals (necessary to derive representative error values on the retrieved parameters) we found it necessary to multiply these errors by a factor of $\sim 15$. Although this may appear to be an alarmingly high factor, we will see later in Fig. \ref{fig:fitspectra} that this leads to error bars on the synthetic $I/F$ reflectivity spectra of only 0.5 -- 1.0 \%, which is perfectly reasonable given the likely accuracy of the absorption coefficients used and also the simplicity of our retrieval scheme. We also decided that this approach was more appropriate than our usual procedure of simply adding a forward modelling error (which here would have been 0.5 -- 1.0\% at all wavelengths and locations) since this would miss the fact that the Minnaert-fitting errors are dependent on both wavelength and latitude.

\begin{table} 
\caption{Five-point Gauss-Lobatto Quadrature Scheme used in this study}
\centering
\begin{tabular}{c l l l}
\hline
Index  & $\mu$ & $\theta$ ($^\circ$) & Weight \\
\hline
1 &  0.1652790  & 80.4866 &  0.3275398 \\
2 &  0.4779249  & 61.4500 &  0.2920427 \\
3 &  0.7387739  & 42.3729 &  0.2248893 \\
4 &  0.9195339  & 23.1420 &  0.1333060 \\
5 &  1.0000000  & 0.00000  &  0.0222222 \\
\hline
\label{tab:quadrature}
\end{tabular}
\end{table}

As with our previous analysis \cite{irwin19}, we modelled the atmosphere of Neptune using 39 layers spaced equally in log pressure between $\sim 10$ and $0.001$ bar. We ran NEMESIS in correlated-k mode and for methane absorption used a methane k-table generated from the band model of \citeA{kark10}. The collision-induced absorption of H$_2$-H$_2$ and H$_2$-He near 825 nm was modelled with the coefficients of \citeA{borysow89a,borysow89b,borysow00}, assuming a thermally-equilibriated ortho:para hydrogen ratio. Rayleigh scattering was included as described in \citeA{irwin19} and the effects of polarization and Raman scattering were again justifiably neglected at these wavelengths. We used the solar spectrum of \citeA{chance10}, smoothed with a triangular line shape of FWHM = 2 nm and took Neptune's distance from the Sun on the date of observation to be 29.94 AU. The reference temperature and mole fraction profile is the same as that used by \citeA{irwin19} and is based on the  `N' profile determined by Voyager-2 radio-occultation measurements \cite{lindal92}, with He:H$_2$ = 0.177 (15:85), including 0.3\% mole fraction of N$_2$. 

\begin{figure} 
\includegraphics[width=\textwidth]{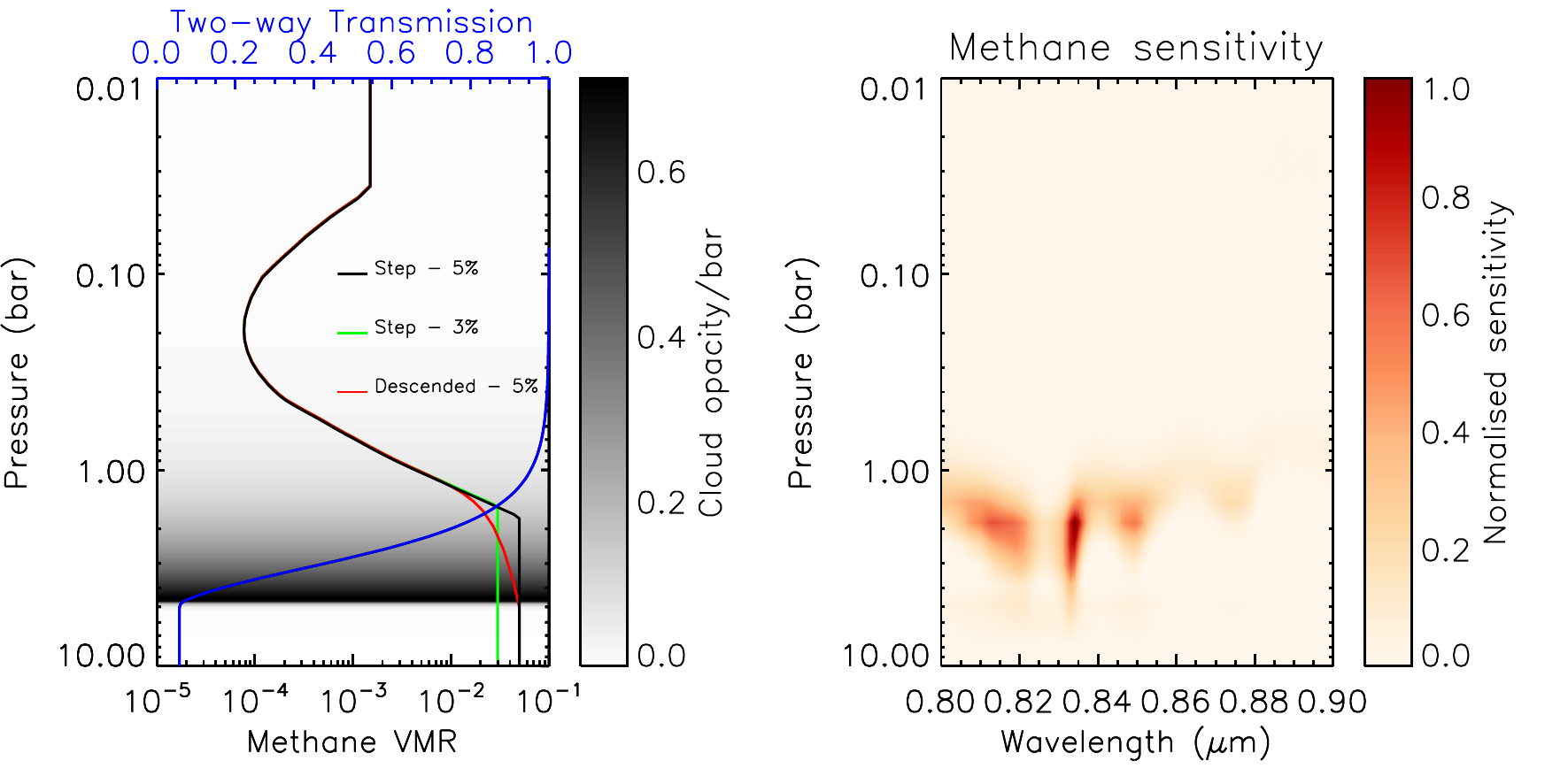}
\caption{Left hand panel: Methane profiles considered in this analysis. Our simple ``step" model is shown for two different values of the `deep' methane mole fraction (3\% and 5\%, respectively) and compared with the ``descended" methane profile of \citeA{sromovsky19} for a deep mole fraction of 5\%, a deep pressure of $P_d = 5$ bar and scaling coefficient, $vx = 3.0$ (see Eq. 3 of \citeA{sromovsky19} for details). The shaded region is the cloud opacity/bar of our nominal equatorial cloud distribution model, which has a base pressure of 4.66 bar. Note that both models have been limited to 100\% relative humidity at all pressures and the stratospheric mole fraction is limited to not exceed $1.5 \times 10^{-3}$. Also plotted for reference is the two-way transmission from space to each level through the main cloud only, showing the cloud to be mostly opaque. The haze distribution has been omitted from this figure for clarity. Right-hand panel: sensitivity of calculated radiances on the abundance of methane at each pressure level, showing that our main sensitivity is from 1--4 bar.} \label{fig:methane_prof}
\end{figure}

For the methane profile, we adopted a simple model with a variable deep mole fraction, limited to 100\% relative humidity above the condensation level and further limited to a maximum stratospheric mole fraction of $1.5\times 10^{-3}$ \cite{lellouch10} as shown in Fig. \ref{fig:methane_prof}. Several authors \cite[e.g.,]{kark11,sromovsky19} have pointed out that such a simple ``step" model is not physically well based for either Neptune or Uranus when extended to great depths and in particular \citeA{tollefson19} notes that such strong deep methane latitudinal gradients would induce humidity winds (additional to the thermal wind equation), which are not seen. Instead, \citeA{sromovsky19} favours a ``descended profile" model, where downwards motion suppresses the methane mole fraction in the 2--4 bar region, but which then recovers to a uniform deep mole fraction at depth. This profile is compared with our ``step" model in Fig. \ref{fig:methane_prof}. Although this profile is smoother and may be more physically plausible than the ``step" model, we do not have the vertical resolution in the MUSE data to be able to discriminate between the two and we also cannot see clearly through the H$_2$S cloud to deeper pressures. This can be seen in Fig. \ref{fig:methane_prof}, where we have also plotted the two-way vertical transmission to space through the cloud only, which shows the fitted cloud to be nearly opaque. In addition, we have also plotted in Fig. \ref{fig:methane_prof} the functional derivatives with respect to methane abundance, i.e., the rate of change of the calculated radiance spectrum with respect to the methane abundance at each level if we were to assume a continuous profile. Here we can see that we are only significantly sensitive to the methane abundance in the 1--4 bar region. In fact, we see that with the MUSE data we are only really sensitive to the column abundance of methane above the H$_2$S cloud and this column abundance will depend on the vertical distribution of both the methane mole fraction and the cloud; since we do not have precise constraints on either we thus have a degeneracy. Hence, in this study we 
decided used the simpler ``step" model for methane, which has the added advantage of returning a mean value for the methane mole fraction in the 2--4 bar region, which is easy to understand, interpret and compare with previous studies. It is worth noting that \citeA{tollefson19}, who were able to probe to slightly deeper pressures than we were, also adopted a simple ``step" model of methane. 

 For clouds/hazes we again adopted the parameterized model used by \citeA{irwin16} to model VLT/SINFONI and Gemini/NIFS H-band observations of Neptune, which was found to provide good limb-darkening/limb-brightening behaviour. In this model particles in the troposphere are modelled with a cloud near the H$_2$S condensation level (and which is thus presumed to be rich in H$_2$S ice \cite{irwin19}), with a variable base pressure ($\sim 3.6$ -- 4.7 bar) and a scale height retrieved as as a fraction of the pressure scale height (called the fractional scale height). Scattering from haze particles is modelled with a second layer with base pressure fixed at $0.03$ bar and fixed fractional scale height of 0.1. Although the base pressure of the stratospheric haze may in reality vary with latitude, we found that the precise pressure level did not  significantly affect the calculated spectra at these wavelengths (since the transmission to space is close to unity at the tropopause level from 800 to 900 nm) and so fixed it to a typically representative value stated. The scattering properties of the cloud were calculated using Mie scattering and a retrievable imaginary refractive index spectrum. For cases where we allow the imaginary refractive index to vary with wavelength (as in our previous report)  we use a Kramers-Kronig analysis to construct the real part of the refractive index spectrum, assuming $n_{real}=1.4$ at 800 nm. Here, however, for simplicity we forced the imaginary refractive indices to be the same at all wavelengths across the 800 -- 900 nm range considered, and hence the real refractive index was fixed to 1.4 over the whole range also. The Mie-calculated phase functions were again approximated with combined Henyey-Greenstein functions for computational simplicity and also to smooth over features peculiar to spherical particles, such as the back-scattering `glory'.
 
 \begin{table} 
\caption{Preset grid of cloud parameterization values used in test retrievals of cloud opacity, cloud fractional scale height, stratospheric haze opacity and cloud-top methane  mole fraction at the equator and $60^\circ$S}
\centering
\begin{tabular}{l l l l}
\hline
Parameter  & Number of values  & Values \\
\hline
Mean cloud radius & 4 & 0.05, 0.1, 0.5, 1.0 $\mu$m \\
Cloud radius variance & 2 & 0.05, 0.3 \\
Cloud $n_{imag}$ & 3 & 0.001, 0.01, 0.1 \\
Mean haze radius & 4 & 0.05, 0.1, 0.5, 1.0 $\mu$m \\
Haze radius variance & 2 & 0.05, 0.3 \\
Haze $n_{imag}$ & 3 & 0.001, 0.01, 0.1 \\
$p_{base}$ & 3 & 3.65, 4.15, 4.66 bar \\
\hline
\label{tab:parameter}
\end{tabular}
\end{table}

\subsection{Retrieval analysis}\label{retrieval_analysis}

\begin{figure} 
\includegraphics[width=\textwidth]{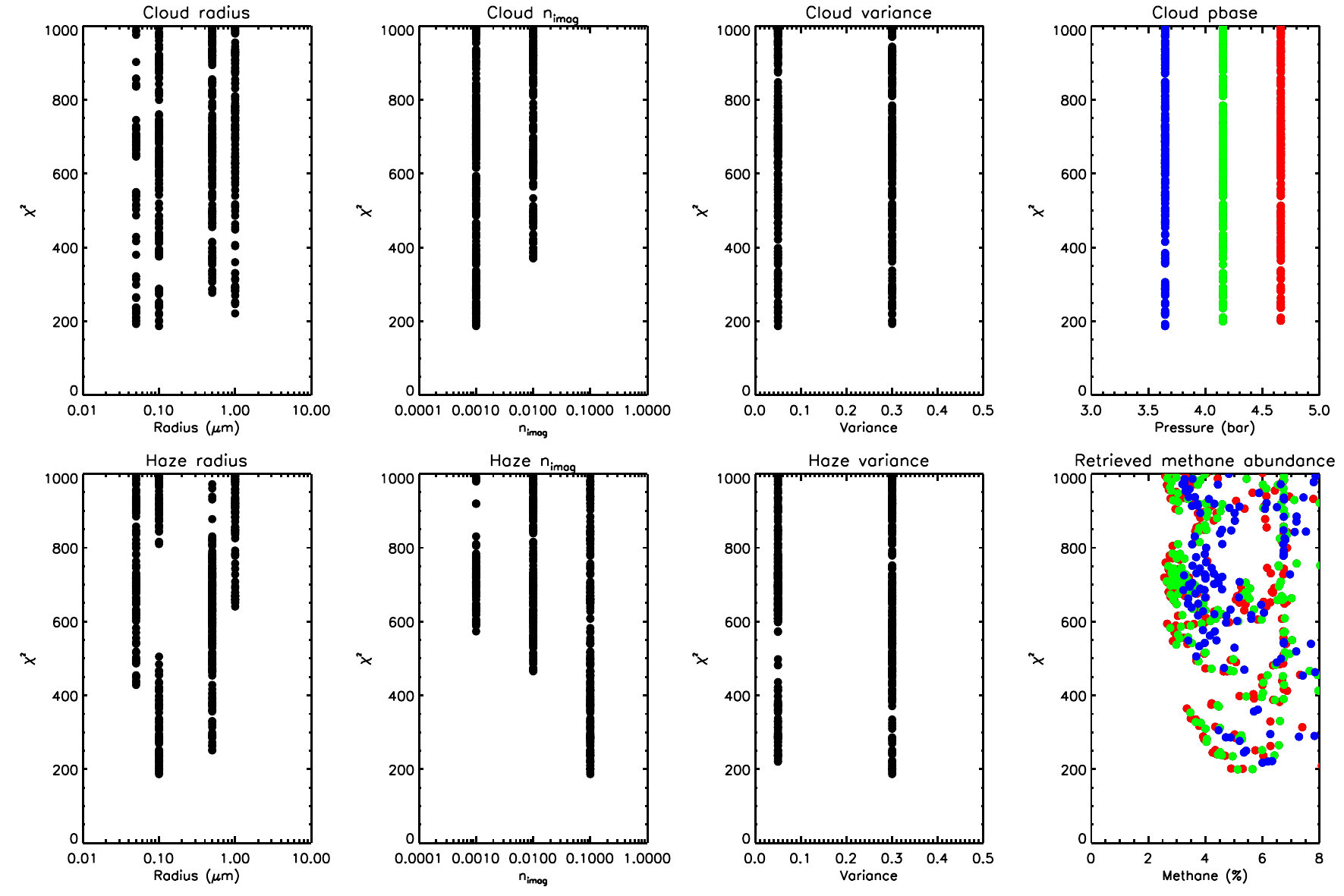}
\caption{Variation of the goodness-of-fit of our retrievals ($\chi^2$) in the equatorial band (variable cloud fractional scale height, stratospheric haze opacity and cloud-top methane  mole fraction) as a function of the fixed grid values of the other cloud and haze properties defined in Table. \ref{tab:parameter}. The first three panels of the top row show the $\chi^2$ values of all the fits for different fixed values of the mean cloud particle radius, cloud particle imaginary refractive index, and variance of the cloud particle radius distribution, while the first three panels of the bottom row show the $\chi^2$ values  for the corresponding haze particle properties. The top right panel shows the $\chi^2$ values for different set values of the cloud base pressure, while the bottom right panel shows the fitted values of the cloud-top methane mole fraction, with the cloud base pressure of each case colour-coded as indicated in the panel above to show that there is no simple correlation between retrieved methane mole fraction and set cloud base pressure. As we are fitting simultaneously to $2 \times 101 = 202$ points in total (i.e., two spectra at $0^\circ$ and $61.45^\circ$ zenith angle, respectively), $\chi^2 =\  \sim200$ indicates a good fit. } \label{fig:equator}
\end{figure}

\begin{figure} 
\includegraphics[width=\textwidth]{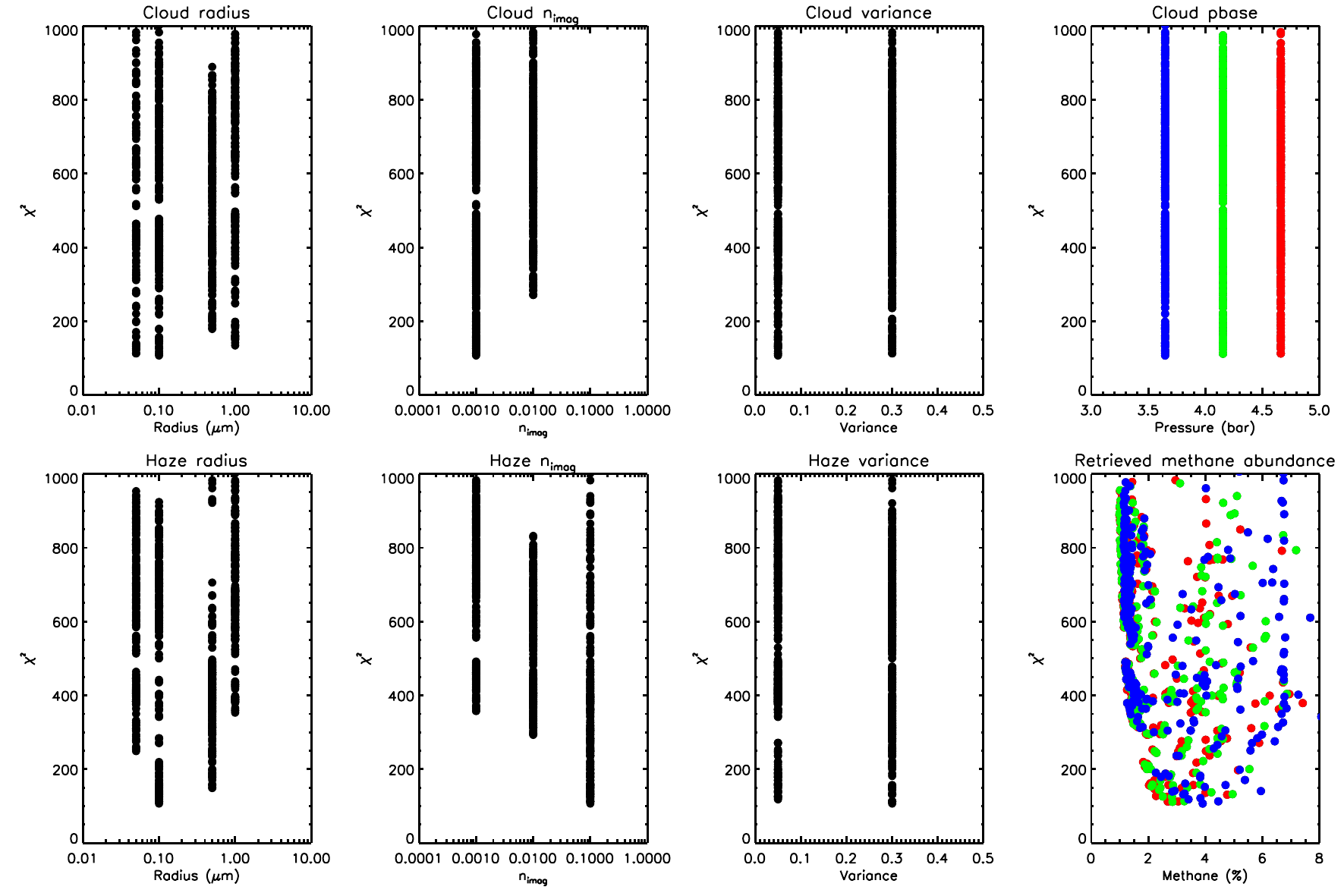}
\caption{As Fig. \ref{fig:equator}, but for the $60^\circ$S latitude band.} \label{fig:60south}
\end{figure}

To `tune' our retrieval model we first concentrated on the latitude bands at the equator and $60^\circ$S. We were aware from our previous study that there was likely to be a high degree of degeneracy in our best-fit solutions with respect to assumed particle sizes and other parameters in the 800 -- 900 nm range. Hence, we first analysed these two latitude bands for a grid of preset values of: 1) mean cloud particle radius; 2) variance of cloud radius distribution; 3) cloud imaginary refractive index; 4) mean haze particle radius; 5) variance of haze radius distribution; 6) haze imaginary refractive index; and 7) cloud base pressure, described in Table~\ref{tab:parameter}. From Table~\ref{tab:parameter} we can see that the number of grid values is $4 \times 2 \times 3 \times 4 \times 2 \times 3 \times 3 = 1728$ setups for each latitude band. For each setup we retrieved simultaneously four variables from the synthetic spectra reconstructed at 0$^\circ$ and 61.45$^\circ$ emission angle: 1) cloud opacity; 2) cloud fractional scale height; 3) haze opacity; and 4) the cloud-top methane mole fraction. As explained earlier, we assumed that the set imaginary refractive indices applied at all wavelengths simultaneously. After fitting we plotted the $\chi^2$ of the fits as a function of the grid parameters, together with the retrieved cloud-top methane mole fraction, which we show in Fig.~\ref{fig:equator} for the equator and Fig.~\ref{fig:60south} for the $60^\circ$S. Here we can see that the goodness of fit 
depends little on the assumed deep pressure of the cloud, nor on the variance of the size distribution of the cloud and haze particles. However, we can see that there is a strong preference for solutions with a haze particle mean radius of 0.05 -- 0.1 $\mu$m and high imaginary refractive index of 0.1. For the cloud, it can be seen that the preference is for a low imaginary refractive index of 0.001. Retrievals where the cloud imaginary refractive index was set to 0.1 had  $\chi^2$ in excess of 1000 and so are not visible in Figs.~\ref{fig:equator} and \ref{fig:60south}, but the constraint on the cloud mean radius is not strong with values of 0.05 -- 0.1 $\mu$m slightly favoured over 1.0 $\mu$m. A cloud mean radius of 0.5 $\mu$m is least favoured. In addition, we checked to see if the limb-darkening curves modelled with our radiative transfer model at all zenith angles were consistent with the Minnaert law and found very good correspondence for zenith angles less than $\sim 70^\circ$ (noted in Fig. \ref{fig:limbcurve}) for schemes using both five and nine zenith angles, adding confidence to our approach.

Although there are a wide range of best-fit $\chi^2$ values it is apparent that the best fits are achieved for a methane cloud-top mole fraction of $\sim$ 4--6\% at the equator and $\sim$ 2--4 \% at $60^\circ$S. However, although we clearly retrieve lower methane mole fractions near the south pole than at the equator it can be seen that there are a wide range of possible cloud solutions that give equally good fits to the data, but rather different methane abundances. Hence, although it would appear that the polar methane cloud-top mole fraction is $\sim 0.5$  times that at the equator we can be less certain of the absolute cloud-top methane mole fraction at the equator and pole.  

\begin{figure} 
\includegraphics[width=\textwidth]{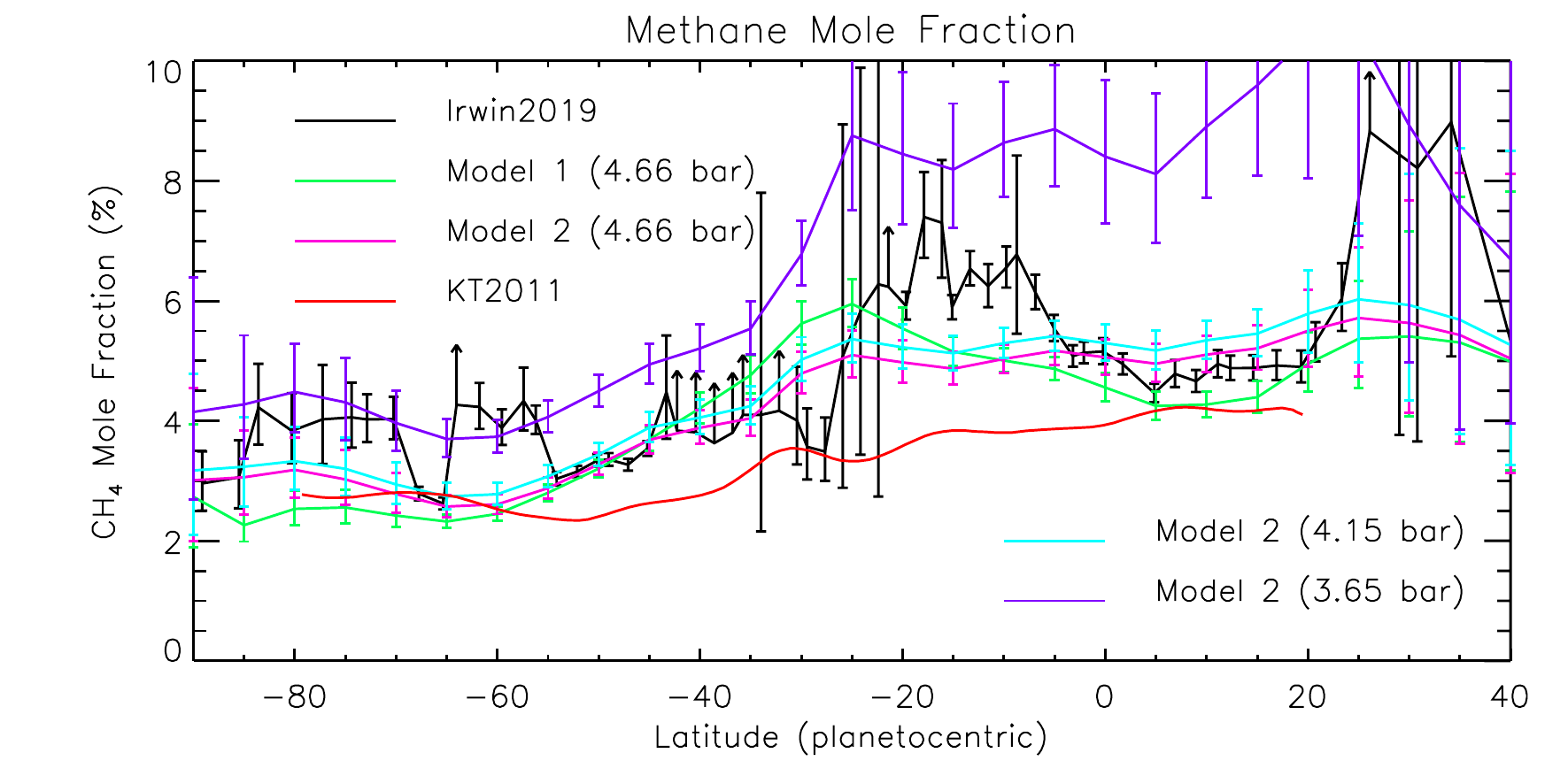}
\caption{Fitted methane mole fractions as a function of latitude. The results from our previous work \cite{irwin19} are shown for reference and compared with our new model: 1) where the imaginary refractive indices of the haze and cloud are fixed to 0.1 and 0.001, respectively; and 2) where the imaginary refractive indices of the haze and cloud are allowed to vary (keeping constant with wavelength). Also shown are the methane mole fractions estimated by \citeA{kark11}, scaled to match our estimates, and recalculations with Model 2 where the base cloud pressure has been reduced to 4.15 and 3.65 bar, respectively. The difference is not large for $p_{base}=4.15$ bar, but for $p_{base}=3.65$ bar it can be seen that the methane retrieval becomes unstable, for the reasons described in the text.} \label{fig:fitnominal}
\end{figure}

Having surveyed the range of cloud properties that best match the observed limb darkening at the equator and $60^\circ$S, we then took one of the best-fit setup cases and applied this to all latitudes. We chose to fix $p_{base} = 4.66$ bar, $r_{cloud} = 0.1$ $\mu$m with 0.05 variance and imaginary refractive index $n_{imag}=0.001$. For the haze we chose to fix  $r_{haze} = 0.1$ $\mu$m with 0.3 variance and imaginary refractive index $n_{imag}=0.1$. We then fitted to the synthetic spectra generated from our fitted Minnaert limb-darkening coefficients for all the latitude bands sampled by the Neptune MUSE observations and fitted once more for 1) cloud opacity; 2) cloud fractional scale height; 3) haze opacity; and 4) the cloud-top methane mole fraction. The resulting fitted methane cloud-top mole fractions as a function of latitude are shown in Fig.~\ref{fig:fitnominal}, where we also show the methane mole fraction variation derived in our previous analysis \cite{irwin19} and that derived by \citeA{kark11}. In Fig. \ref{fig:fitnominal} we can see that our derived latitudinal methane distribution for our default model (indicated as Model 1) varies much more smoothly with latitude than our previous analysis \cite{irwin19} and has more smoothly-varying error bars. In addition, it can be seen that our new retrieved methane variation more closely resembles that determined by \citeA{kark11}. The greatest discrepancy occurs at 20 -- 40$^\circ$S and we found here that our fits had the highest $\chi^2/n$ values. To introduce additional flexibility into our model, we ran our retrievals a second time, but additionally allowed the model to vary the imaginary refractive indices of the cloud and haze particles (Model 2), where the imaginary refractive indices were still assumed to be wavelength invariant. It can be seen that Model 2 retrieves lower methane mole fractions at 20 -- 40$^\circ$S and even more closely resembles the results of \citeA{kark11}. 

\begin{figure} 
\includegraphics[width=\textwidth]{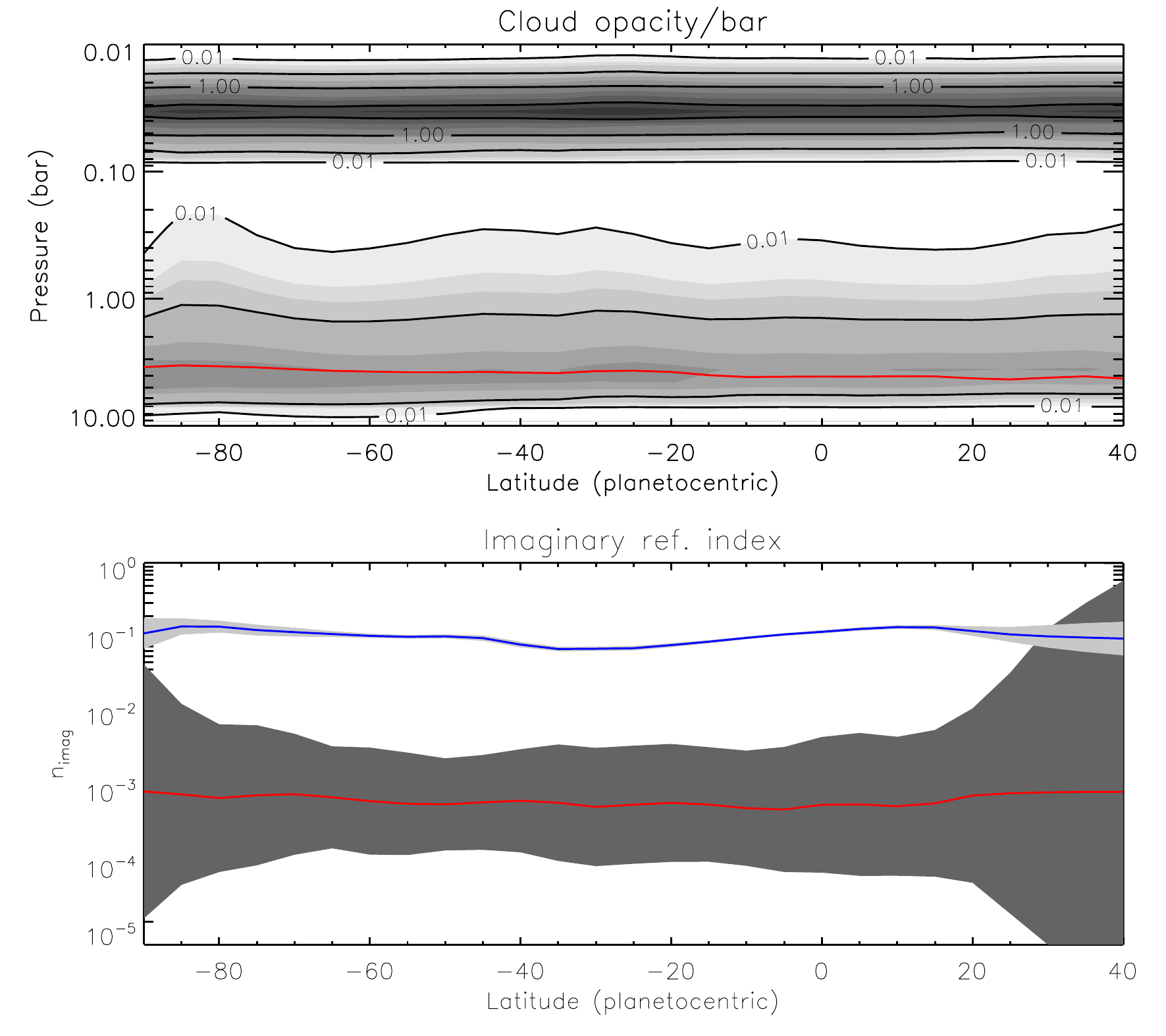}
\caption{Fitted latitudinal variation of cloud opacity/bar  (at 800 nm) and imaginary refractive indices of the cloud and haze particles. The top panel shows a contour plot of the fitted cloud opacity/bar profiles (darker regions indicate greater cloud density), while the bottom panel shows the latitudinal variation of the retrieved imaginary refractive indices of the cloud particles (red) and haze particles (blue), together with error range (grey).  For the imaginary refractive indices it can be seen that this is poorly constrained for the cloud (large error bars) and we just need the particles to be highly scattering. However, the imaginary refractive index of the haze is well constrained, and shows these particles are more scattering at 20--40$^\circ$S. The red line in the top plot indicates the cloud top pressure (i.e., level where overlying cloud opacity at 800 nm is unity). The cloud contour map indicates the main cloud top to lie at similar pressure levels at all latitudes and has a cloud-top pressure of $\sim$ 3--4 bar. We can also see a very slight increase in stratospheric haze opacity at 20 -- 40$^\circ$S, associated with the cloudy zone and then clearing slightly towards the north and south.} \label{fig:fitrefindex}
\end{figure}

Fig. \ref{fig:fitrefindex} shows the latitudinal variation of the wavelength-invariant values of $n_{imag}$ retrieved by Model 2 for the cloud and haze, and also shows the retrieved 2-D (i.e., latitude -- altitude) cloud structure. We can see that  $n_{imag}$ for the cloud is poorly constrained, but that of the haze is well estimated and it would appear that to best match the observations at 20 -- 40$^\circ$S the haze particles are required to have slightly lower $n_{imag}$ values than those found at other latitudes. This is easily understood looking at Fig. \ref{fig:neptune_recon} where we can see that this latitude has numerous bright, high, discrete clouds. Although we masked the observations to focus the retrievals on the background smooth latitudinal variation, to mask completely the brighter clouds at these latitudes would have left us with no data to analyse at all (Fig.\ref{fig:neptune_cut}). Hence, we would expect Model 2, which allows the cloud/haze particle reflectivity to vary, to better incorporate the additional reflectivity from these upper tropospheric methane clouds and so fit the observations more accurately and also retrieve a more reliable latitudinal variation in cloud-top methane mole fraction. Please note that the cloud opacity plot in Fig.~\ref{fig:fitrefindex} shows opacity below the 4.66-bar cloud base pressure for two reasons: 1) we assume the opacity to diminish with a scale height of 1 km below the condensation level rather than cutting off sharply; and 2) we show here the opacity in the 39 model atmospheric layers, which are split equally between $\sim 10$ and $0.001$ bar and so do not coincide exactly with the base pressures of the cloud and haze. 

\begin{figure} 
\includegraphics[width=\textwidth]{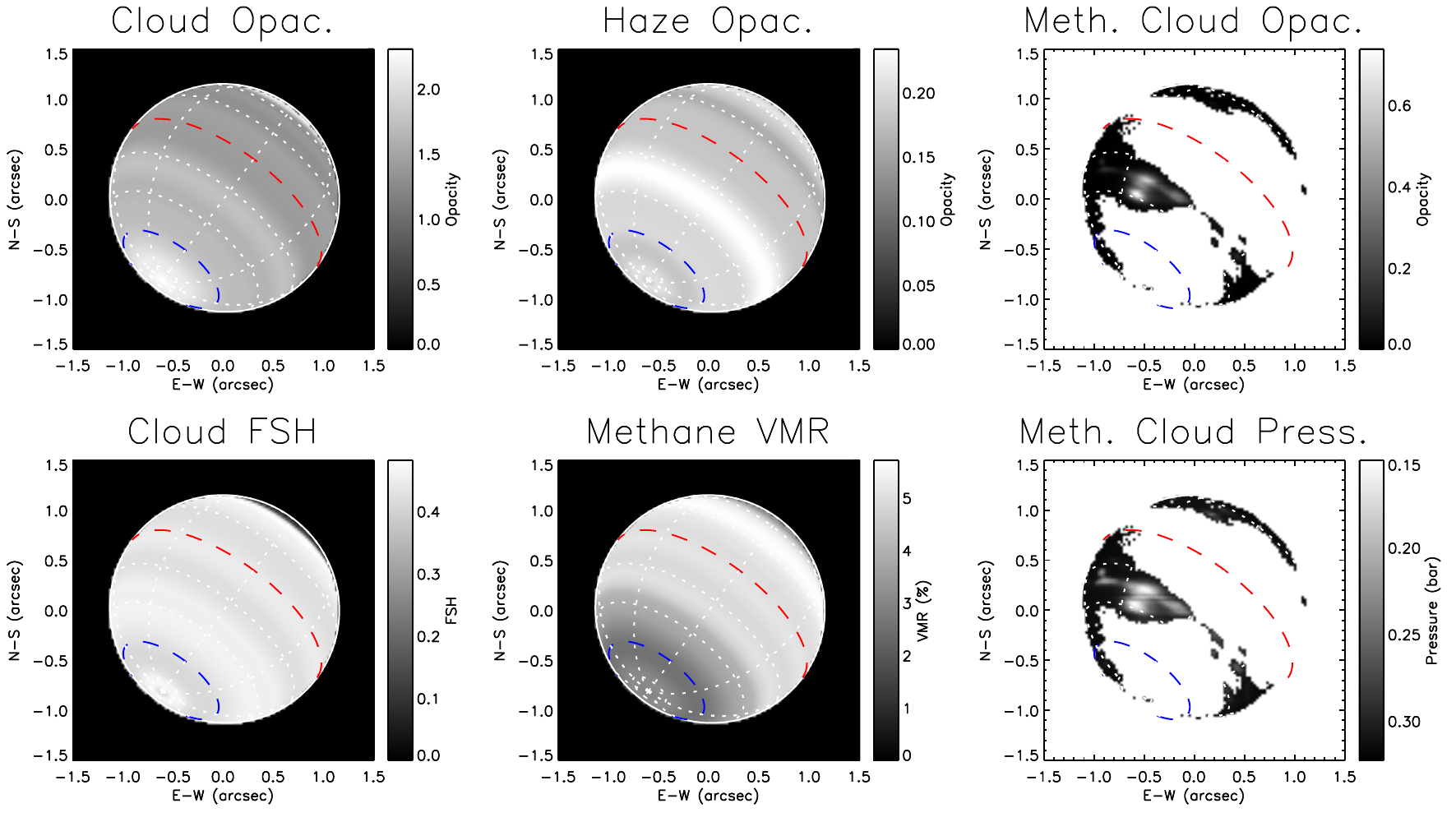}
\caption{Fitted tropospheric cloud opacity (Cloud Opac.), tropospheric cloud fractional scale height (Cloud FSH), stratospheric cloud opacity (Haze Opac.) and cloud-top methane  mole fraction (Methane VMR) projected on to the Neptune's disc as seen by VLT/MUSE. Also plotted are the opacity (Meth. Cloud Opac.) and mean pressure level (Meth. Cloud Press.) of additional methane clouds used to fit the discrete cloud regions. The methane mole fraction plot clearly shows lower values of methane polewards of $30-40^\circ$S and an apparent possible local minimum at $60^\circ$S. } \label{fig:retA}
\end{figure}

In addition to providing a better constrained retrieval of cloud-top methane mole fraction, showing its mole fraction to decrease from equator to south pole, our new retrieval scheme appears to detect noticeably lower mole fractions of methane near 60$^\circ$S. This is more easily seen in Fig. \ref{fig:retA}, which shows the spatial variation of tropospheric cloud opacity, tropospheric cloud fractional scale height, stratospheric cloud opacity and cloud-top methane mole fraction projected onto the disc of Neptune as seen by VLT/MUSE. It is difficult to be certain if this is a real feature as we have much less geometrical coverage of the limb-darkening curves as we approach the south pole. If it is a real feature then it is possible it might perhaps be related to the South Polar Feature (SPF), e.g., \citeA{tollefson19}.  We will return to this question in the next section.

\begin{figure} 
\includegraphics[width=\textwidth]{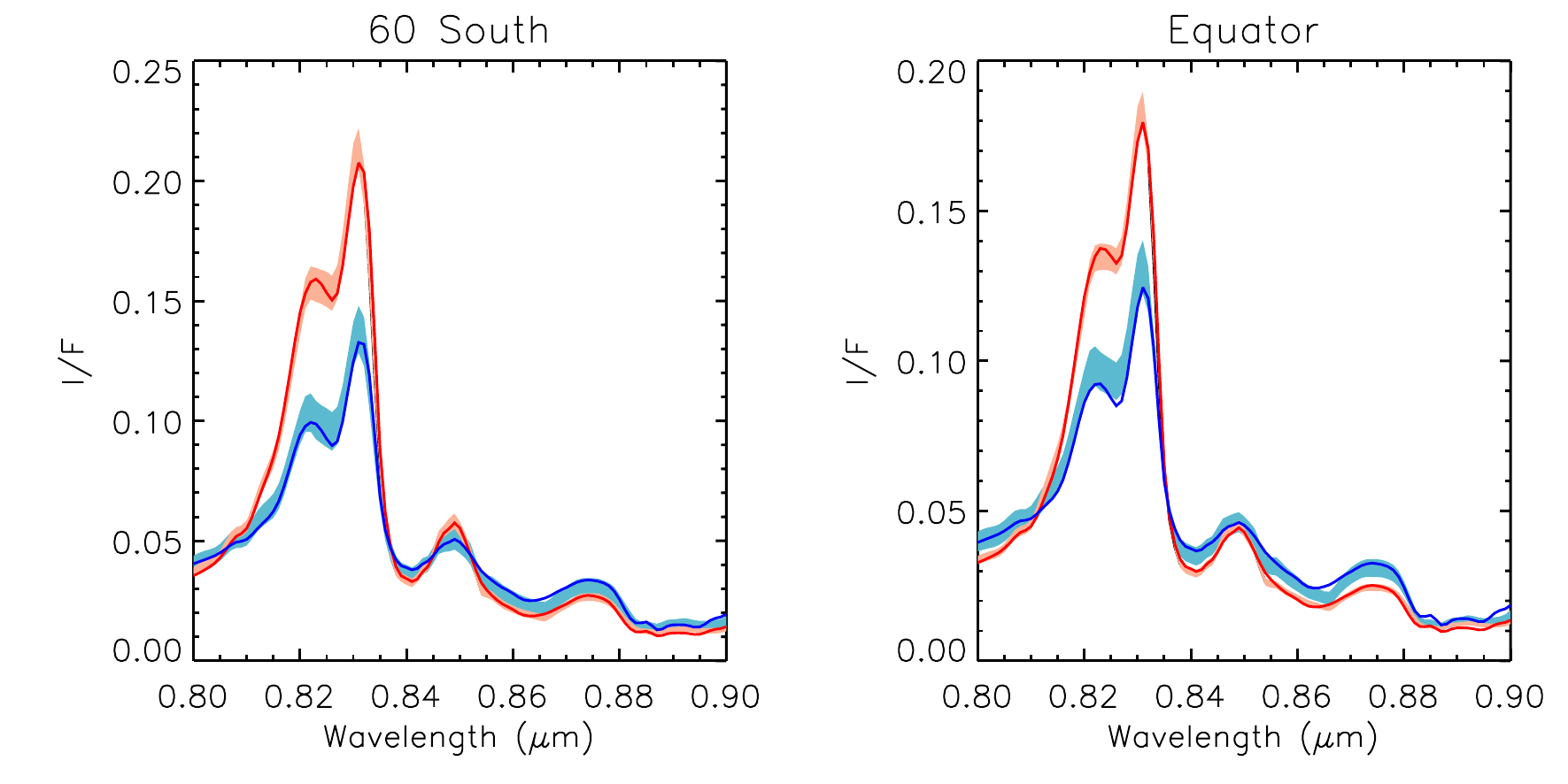}
\caption{Spectra fitted by NEMESIS to synthetic observations at $0^\circ$ (red) and $61.45^\circ$ (blue) zenith angle generated from the Minnaert limb-darkening analysis at $60^\circ$S and the equator. The Minnaert-modelled synthetic observed spectra and error limits (as described in the text) are shown as the light shaded regions, while the spectra fitted by our retrieval model are shown as the solid, darker lines. } \label{fig:fitspectra}
\end{figure}

Finally, in Fig.\ref{fig:fitspectra} we show the fitted spectra from Model 2 at $0^\circ$ and $61.45^\circ$ zenith angle, compared with the synthetic `observed' spectra at the equator and $60^\circ$S. The error bars of the synthetic observations are shown as the lighter colour-shaded regions and it is apparent here why the errors in the synthetic observations had to be inflated to enable the retrieval model to fit to an accuracy of $\chi^2/n \sim 1$: even when inflated the reflectivity errors are still small ($\sim 0.5$ \%) compared with the likely accuracy of the gaseous absorption coefficients used and the simplicity of our cloud parameterization scheme.

\subsection{Comparison with previous retrievals}

The difference between our new methane retrievals and our previous estimates \cite{irwin19} are for some locations greater than 3-$\sigma$, although it should be remembered that these are random errors only, and do not account for systematic errors arising from differences in the assumed methane/cloud models. These differences mostly occur in the cloud belt near 20 -- 40$^\circ$, where we have unaccounted upper tropospheric methane ice clouds, but we wondered whether there might be other effects that might explain the sharper latitudinal changes in methane abundance and estimated errors of \cite{irwin19}. Our previous retrievals assumed a base cloud pressure of $p_{base} = 4.23$ bar, rather than $p_{base} = 4.66$ bar as we assumed here. Hence, we re-ran our retrievals using the two other base pressures listed in Table \ref{tab:parameter} of 4.15 and 3.65 bar, respectively. 

The results for Model 2 (where we also fit for $n_{imag}$ of both cloud and haze) for all three cloud base pressures are shown in Fig. \ref{fig:fitnominal}. As can be seen, the results for $p_{base} = 4.15$ bar are very similar to those for $p_{base} = 4.66$ bar, but those for $p_{base} = 3.65$ are very different and appear, in terms of scatter and inflated error bars, more like our previous results \cite{irwin19}. We believe this to be caused by an artefact of our retrieval model, where we have assumed a single cloud with fixed base pressure and variable scale height combined with our simple  ``step" methane model. For Model 2 with $p_{base} = 4.66$ bar it can be seen in Fig. \ref{fig:fitrefindex} that the retrieved level of unit cloud optical depth is in the range 3--4 bar, depending on latitude, comfortably greater than the methane condensation pressure. However, when the cloud base pressure is lowered to $p_{base} = 3.65$ bar the cloud opacity has to be greater at lower pressure levels in order to give enough overall reflectivity. This pushes the level of unit optical depth to lower pressures and, depending on the deep methane abundance, can at some latitudes  become similar to the methane condensation pressure. In such circumstances the sensitivity of the calculated reflectivity to the deep methane mole fraction is reduced and the retrieved mole fraction may need to be greatly increased (and have greater error bars) to give enough methane absorption, exactly as we see.  The retrieved pressure levels of unit optical depth from \citeA{irwin19} are shown in Fig. 9 of that paper to be in the range 1.8--3 bar, which is indeed rather close to the methane condensation pressure level and so would unfortunately have suffered from this same systematic artefact. However, \citeA{irwin19} assumed $p_{base} = 4.23$ bar, a value that gave consistent results in our new retrievals, which indicates that there must be an additional difference between the two analyses that caused \citeA{irwin19} to retrieve unit optical depth values near the methane condensation level.  We have identified this difference to be that rather than using \textit{a priori} values of $n_{imag} = 0.001$ and 0.1 for the cloud and haze respectively, as used in this study, \citeA{irwin19} assumed the imaginary refractive index spectra to be fixed at all latitudes to those retrieved from their limb-darkening analysis at 5 -- 10$^\circ$S. Figure 6 of \citeA{irwin19} shows that the haze particles were found at the equator to be rather dark ($n_{imag}$ in the range 0.076 to 0.159, depending on wavelength). These darker, wavelength-dependent haze particles, combined with the extended wavelength region of 770 -- 930 nm (compared with 800 -- 900 nm considered here) led to larger retrieved haze opacities and consequently required larger cloud opacities to match the peak reflectivity at continuum wavelengths. This then led to the retrieved unit cloud optical depth levels approaching the methane condensation pressure. 

\begin{figure} 
\includegraphics[width=\textwidth]{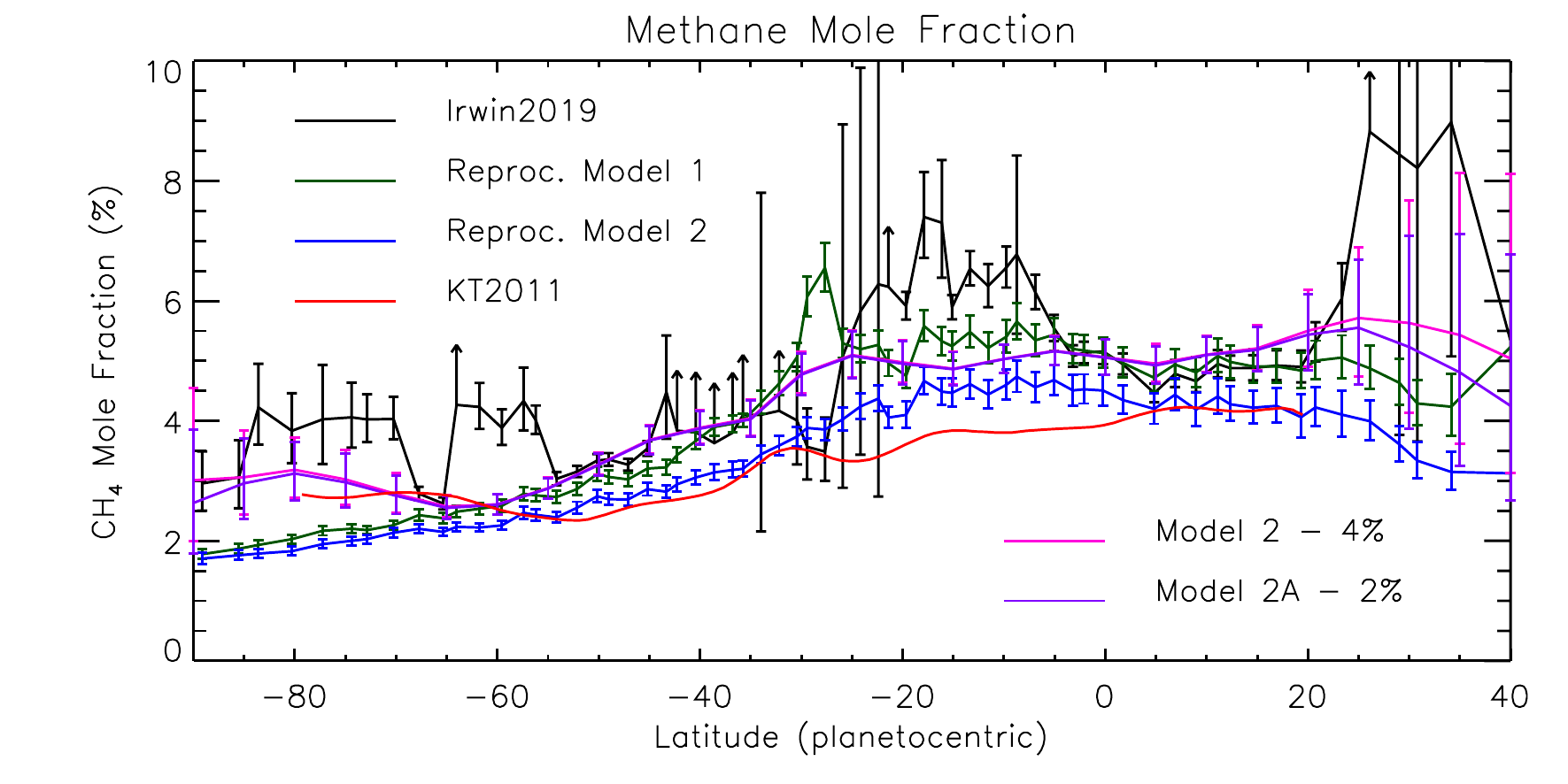}
\caption{As Fig.~\ref{fig:fitnominal}, but comparing the results from our previous work \cite{irwin19} with the results of a reprocessing of the central meridian spectra considered by \citeA{irwin19}, where the cloud/haze scattering properties have been replaced by those used in our new analysis and either fixed at all latitudes (Reproc. Model 1), or allowed to vary with latitude (Reproc. Model 2). In addition, we show the results of our limb-darkening analysis with Model 2 (previously shown in Fig.~\ref{fig:fitnominal} and which has an \textit{a priori} methane mole fraction of 4\%) and a revised version of Model 2, where the \textit{a priori} methane mole fraction has been reduced to 2\% (Model 2A).}  \label{fig:fitmodnominal}
\end{figure}

To demonstrate this effect we repeated the analysis of the central median observations of \citeA{irwin19}, using exactly the same setup as used in this previous study, but substituting the \textit{a priori} cloud scattering properties to be those used by our new Model 1 (fixed $n_{imag}$) 
and Model 2 (variable $n_{imag}$). 
Our re-fitted methane mole fractions are shown in Fig. \ref{fig:fitmodnominal}. Here, we can see that when reprocessed in this way the set of spectra along Neptune's meridian return a latitudinal variation in deep methane mole fraction that is much more consistent with our new analysis and also with the HST/STIS determinations \cite{kark11}. The apparently small retrieved errors of the reanalysis towards the south pole should be viewed with caution. Such latitudes are only seen over a very small range of zenith angles and so we cannot say anything about the limb-darkening here. Hence, we are much more dependent at these latitudes on the assumed cloud and methane profile, and so our methane estimates are prone to larger systematic error.  In addition, for the central meridional analysis we used the MUSE pipeline radiance errors (scaled to give $\chi^2/n \sim 1$ for fits to spectra near the equator), which are smaller near the pole giving smaller apparent methane mole fraction errors.  In contrast our new limb-darkening analysis has fewer points to define the limb-darkening and so assigns larger error bars to the reconstructed spectra near the pole. Hence, at these latitudes the retrieved deep mole fraction is retrieved with larger error. 

Returning to the question of the apparent minimum of methane at 60$^\circ$S in our new analysis, with larger retrieval errors towards the south pole the solution might be expected to partially relax back to the \textit{a priori} value of ($4\pm 4$)\%\footnote{Note that this parameter is treated logarithmically within NEMESIS, and hence it is the fractional error, i.e., 1.0, that is used in the covariance matrix.}.  However, when we repeated the retrievals using a lower methane mole fraction of ($2\pm 2$)\% (i.e., same fractional error), the same latitudinal behaviour was determined as can be seen in Fig. \ref{fig:fitmodnominal} (Model 2A), so this cannot be the cause. Instead, we believe this apparent methane feature may arise from the limited range of zenith angles sampled to fit the Minnaert parameters at these latitudes, since they will appear only at higher zenith angles and our analysis further omits observations with $\mu > 0.3$ ($\mu$ is the cosine of the zenith angle), to avoid locations too near the disc edge. Figure \ref{fig:spec_contour} shows (bottom right panel) that at these latitudes the Minnaert limb-darkening parameter, $k$, appears to tend to $\sim 0.5$ at $\sim 80^\circ$S at all methane-absorbing wavelengths, but there is no clear difference in the appearance of Neptune at this latitude at any wavelength. Hence, we believe this effect to be a geometrical artefact of our limb-darkening analysis, in the same way that the central meridional analysis shows a continuing decrease towards the pole as we view the locations at higher and higher zenith angle. Only observations recorded with even higher spatial resolution would be able to better constrain the latitudinal variability of methane at such high latitudes. In the meantime, our new determinations of methane mole fraction at polar latitudes have larger retrieval errors properly indicating this greater uncertainty.  

Finally, we return to the question of assumed methane and cloud profile parameterizations and the absolute accuracy of our methane retrievals. As noted earlier, with spectral observations such as these in this wavelength region what we are actually sensitive to is the column abundance of methane above the cloud top. Sadly, the vertical resolution of such nadir/near-nadir observations cannot physically be less than $\sim$ one scale height, which means that it is very difficult to discriminate between the ``step" methane model used here and more sophisticated models such as the ``descended profile" model favoured by, e.g., \citeA{kark11} and \citeA{sromovsky19}. This is especially the case when considering that we do not have a good \textit{ab initio} model for the vertical cloud structure either. It is apparent that for models with higher cloud opacity at lower pressures, the mole fraction of methane will need to be higher to give the same column abundance and so it can be seen that there exists a wide range of possible cloud and methane vertical distribution models that could fit our observations equally well and give the same methane column abundance for a given latitude. For the nominal Model 2 retrievals presented here, with cloud base at 4.66 bar we see a clear reduction in the column abundance of methane towards the pole, which we interpret here in terms of a deep mole fraction varying from ($5.1 \pm 0.3$)\% at the equator to ($2.6 \pm 0.2$)\% at $60^\circ$S, i.e., a reduction factor of $1.9 \pm 0.2$. However, in terms of the latitudinal dependence of the mole fraction of methane this depends on the methane profile, cloud profile and the reference pressure level and so the cloud-top methane mole fraction could conceivably vary by as much as $\sim \pm 1$\%. Hence, here we estimate the equatorial deep mole fraction of methane to be in the range 4--6\%, with the abundance at polar latitudes reduced from those at equatorial latitudes by a factor of $1.9 \pm 0.2$.

\subsection{Extension to discrete cloud retrievals}

Having greatly improved our fit to the background atmospheric state, we wondered if it might be possible to retrieve, in addition, the cloud profiles for the discrete cloud regions masked out in our analysis so far. Discrete clouds such as these are known to exist at pressures from 0.5 -- 0.1 bar \cite[e.g.]{irwin11} and as such are almost certainly clouds of methane ice. It can be seen from Fig. \ref{fig:neptune_recon} that in our observations these clouds are mostly restricted to latitudes 30 -- 40$^\circ$S and are of highly variable reflectivity and only cover a small range of central meridian longitudes. Hence, our Minnaert limb-darkening analysis, which assumes that the clouds at a particular latitude are zonally-symmetric and do not vary with central meridian longitude, was not appropriate for analysing these discrete clouds. Instead, assuming that the cloud properties of the background tropospheric and stratospheric clouds would not be different from their zonally-retrieved Minnaert values, we looked to see what opacity and pressure level additional discrete methane ice clouds might need to have to best match the observed spectra in the previously masked pixels. We assumed that the methane ice particles at such low pressures were likely small and assumed a size distribution with mean radius 0.1 $\mu$m and variance 0.3. Methane ice is highly scattering at short wavelengths \cite{martonchick94,grundy02} and so the complex refractive index was set to $1.3 + 0.00001i$, with scattering properties calculated via Mie theory and phase functions again approximated by combined Henyey-Greenstein functions. For vertical location, we assumed that the cloud had a Gaussian dependence of specific density (particles/gram) with altitude and had a variable peak pressure and opacity. The \textit{a priori} pressure level of the opacity peak was set to 0.3 bar. We reanalysed the areas that had previously been masked and extended the area slightly to capture some of the thinner discrete clouds seen (Fig. \ref{fig:neptune_cut}). Then, for each pixel in this extended area, we set the tropospheric cloud, stratospheric haze and cloud-top methane mole fraction to that determined from the zonally-averaged Minnaert analysis for that latitude and retrieved the opacity and peak pressure of an additional methane ice cloud. 

Our retrieved methane ice cloud properties can be seen in Fig. \ref{fig:retA}, where we retrieve opacities of up to 0.75. Although there are clearly some regions of thick methane ice cloud, the median value of this additional opacity is found to be only 0.0063 and so it only makes a significant difference to the observed radiances in the small discrete regions seen. Figure \ref{fig:retA} also shows an apparent variation in mean pressure of these discrete clouds, remaining near the \textit{a priori} pressure of $\sim 0.3$ bar for the thinnest clouds, but reducing to as low as 0.15 bar for the thickest clouds. However, we find this pressure variation to be insignificant compared with the retrieval errors. Reviewing the two-way transmission-to-space within the 800 -- 900 nm wavelength region examined here, we find that we are only weakly sensitive to the actual pressure level of detached methane clouds in the 0.5 -- 0.1 bar region. Our chosen wavelength band includes the strong methane absorption band at 887 nm, but even here the two-way transmission to space only reduces to 0.5 at the 0.35 bar level for nominal cloud/haze conditions. To really probe the altitudes of such clouds we need to use observations in the much stronger methane bands at 1.7 $\mu$m in the H-band, as has been done by numerous previous authors \cite[e.g.,]{irwin11,irwin16,luszcz16}, but which is not observable by MUSE. We examined the raw MUSE observations near 727 nm and 887 nm, but could not see any clear difference in the brightness of the discrete clouds with wavelength for either weak or bright detached clouds. Hence, all we can really say with the MUSE observations is that the discrete clouds (for all opacities) must lie somewhere at pressures less than $\sim 0.4$ bar. 

\begin{figure} 
\includegraphics[width=\textwidth]{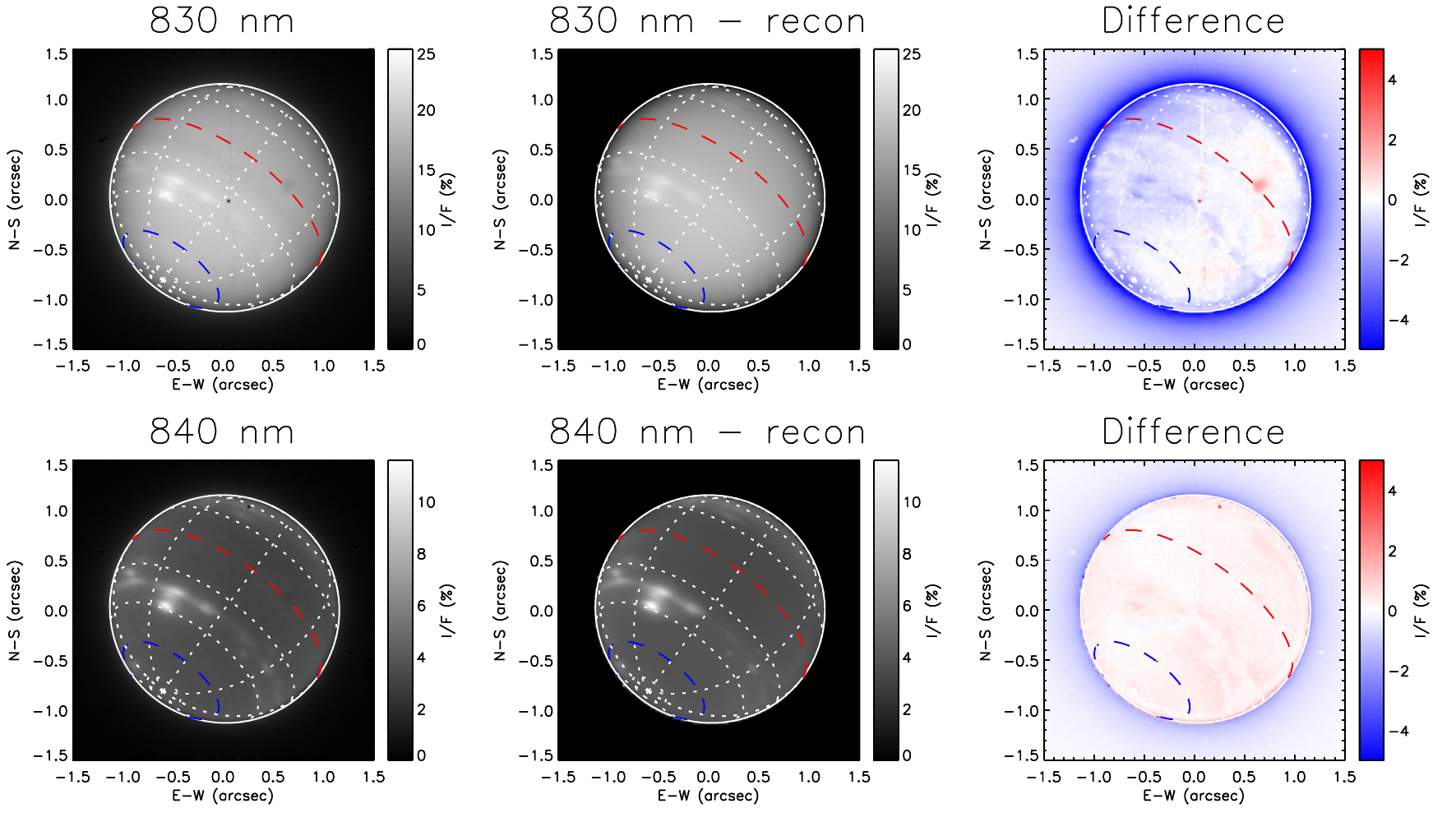}
\caption{As Fig. \ref{fig:neptune_recon}, but here the reconstructed images are generated from NEMESIS forward-modelling calculations from the fitted cloud and methane profiles, including fitting of discrete cloud regions using an additional thin methane ice cloud layer. As can be seen the residuals are reduced greatly and are small at all locations on the planet's disc.} \label{fig:neptune_recon2}
\end{figure}

\begin{figure} 
\includegraphics[width=0.5\textwidth]{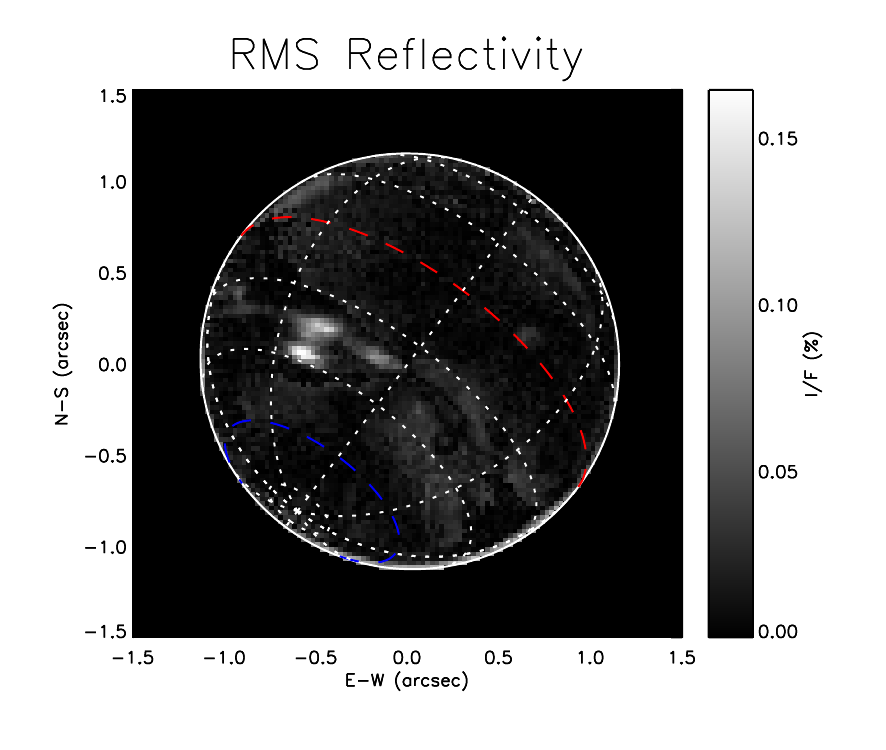}
\caption{Root-mean-square reflectivity (I/F) differences of our final fit to the observations of Neptune over the entire visible disc from 800 -- 900 nm.} \label{fig:rms}
\end{figure}

 With the addition of discrete methane clouds our forward-modelled reconstructed images of Neptune at 830 and 840 nm are compared with the MUSE observations in Fig. \ref{fig:neptune_recon2}. Comparing with Fig. \ref{fig:neptune_recon} it can be seen that we achieve a very good fit at all locations on Neptune's disc. Although we have only shown the fit at two wavelengths here, the fit was found to be very good for all wavelengths in the 800 -- 900 nm range and Fig. \ref{fig:rms} shows the root-mean-square (RMS) reflectivity differences of our fits at all wavelengths, showing that we match the observed spectra to an RMS of typically  0.05\%, increasing to only 0.15\% in the brightest methane ice clouds. At these locations it may be that model is struggling with the fact that the stratospheric haze opacity was set and fixed to that derived from zonally-averaged fits where the discrete clouds had not been entirely masked.

\section{Conclusions}\label{conclude}

In this work, we have reanalysed VLT/MUSE-NFM observations of Neptune, made in June 2018 \cite{irwin19}, with a Minnaert limb-darkening analysis recently developed for Jupiter studies \cite{perez20}. We find that the new scheme allows us to use the observations much more effectively than simply analysing along the central meridian, as we had previously done, as it also accounts for the observed limb-darkening/limb-brightening at different wavelengths and latitudes. Once having fitted the general latitudinal variation of cloud and haze with this analysis we are then able to fit for the properties of discrete methane clouds seen in our observations, allowing us to fit all locations on the visible disc to a reflectivity (I/F) precision of 0.5 -- 1.0\% (RMS $< 0.15$\%). Our main conclusions are:
\begin{itemize}
    \item We find that we are able to fit the background reflectivity spectrum of Neptune from 800 -- 900 nm with a simple two-cloud zonally symmetric model comprising a deep cloud based at 4.66 bar, with variable fractional scale height and a layer of stratospheric haze based near 0.03 bar. 
    \item The cloud-top mole fraction of methane at 2--4 bar (i.e., above the H$_2$S cloud, based at 4.66 bar) is found to decrease by a factor of $ 1.9\pm 0.2$ from equator to pole, from 4--6\% at the equator to 2--4\% at the south pole.
    \item While this latitudinal decrease in methane mole fraction is well defined, the absolute mole fractions at different latitudes depends on the precise choice of cloud parameterization (and indeed methane parameterization), for which a wide range of setups give similar goodness of fit.   
    \item The previous retrievals along the central meridian of these data reported by \citeA{irwin19} appear to have suffered from an unfortunate retrieval artefact at some locations due to a less sophisticated incorporation of limb-darkening. Our new methane retrievals are more robust, vary more smoothly with latitude, and give a clearer and conservative estimate of the likely error limits. 
    \item The opacity of both the tropospheric cloud and stratospheric haze is found to be maximum at 20 -- $40^\circ$S and 20 -- $40^\circ$N, which are also the latitudes of the discrete methane clouds seen. While this may be a real feature, it may also be that the limb-darkening curves analysed by our Minnaert scheme were contaminated by thin discrete clouds at these latitudes. 
    \item Adding localised methane clouds to our zonal model allows us to additionally retrieve the properties of the discrete cloud locations; we find these clouds must lie at pressures less than 0.4 bar and have opacities of up to 0.75.
    \item The latitudinal variation of cloud-top methane mole fraction with latitude observed here in 2018 seems little changed from that seen by HST/STIS in 2003 \cite{kark11}, indicating that this apparent latitudinal distribution of cloud-top methane has not varied significantly in the intervening fifteen years.
\end{itemize}

Having now developed a scheme that matches the observed spectra of Neptune from 800 -- 900 nm, the next step will be to reproduce the appearance of Neptune at other wavelengths to see if our fitted cloud/methane model is more generally applicable. There will be two main challenges to this:

\begin{enumerate}
    \item Firstly, extending to shorter wavelengths the contribution of Rayleigh scattering becomes more important, which limits our ability to see to deeper cloud layers. However, more importantly for the MUSE-NFM observations is the fact that the full-width-half-maximum (FWHM) of the point-spread-function (PSF) becomes significantly worse. It can be seen in Fig. \ref{fig:neptune_recon} that even at 800--900 nm, there is a considerable residual off-disc signal due to the PSF. We could have tried to model this in our fitting procedure, but concluded that it was simpler to limit ourselves to locations not too near the disc edge in the Minnaert analysis. However, this would become more difficult to justify at shorter wavelengths. Instead, in future work we hope to take our existing model, calculate the appearance at shorter wavelengths, and convolve with a PSF model, which still needs to be developed. By thus simulating the shortwave observations we will be able to see if we can discount more of the possible cloud/haze paramaterisation setups, which will help refine our methane retrievals.
    \item Extending to longer wavelengths, the contribution of Rayleigh scattering becomes less and the increasing strength of the methane absorption bands means that it is possible to probe the vertical extent and location of clouds more precisely. We would need to extend to J, H and K-band observations to see which of our cloud/haze setups can be discounted, using the fact that the opacity of small cloud particles will fall more quickly with wavelength than large particles. However, this will mean trying to analyse observations taken at different apparitions and with different instruments, which will mean that the cloud distribution will not be the same and that systematic errors in the photometric calibration and PSF characterisation will arise. In addition, the complex refractive indices of the particles  will be different from those we have derived at these wavelengths.   
\end{enumerate}

Although these difficulties are not insurmountable, they will require careful evaluation and effort to overcome, which is why we leave them as future work. However, it is clear from this study that splitting the atmosphere of Neptune into a background zonal part (that can be fitted with a the Minneart limb-darkening scheme) and an additional spatially-varying part due to discrete methane clouds greatly simplifies the retrieval process and allows us to efficiently reconstruct the entire 3-D structure of Neptune's clouds and methane cloud-top mole fraction. Our model is then simultaneously consistent with the observations at all wavelengths under consideration and all locations on Neptune's disc. The approach could also be applied to observations of Neptune at other wavelengths and also Uranus and Saturn, which to a first approximation appear zonally symmetric with additional discrete cloud features at visible/near-infrared wavelengths. It could also be further applied to Jupiter observations, building upon the work of \citeA{perez20}, at wavelengths and latitudes that appear relatively homogeneous.  

\acknowledgments
We are grateful to the United Kingdom Science and Technology Facilities Council for funding this research. Glenn Orton was supported by NASA funding to the Jet Propulsion Laboratory, California Institute of Technology. Leigh Fletcher was supported by a Royal Society Research Fellowship and European Research Council Consolidator Grant (under the European Union's Horizon 2020 research and innovation programme, grant agreement No 723890) at the University of Leicester.  The observations reported in this paper have the ESO ID: 60.A-9100(K).


%
%

\bibliography{NeptuneMUSE}

\begin{thebibliography}{}

\bibitem [\protect \citeauthoryear {%
{Bacon}%
\ \protect \BOthers {.}}{%
{Bacon}%
\ \protect \BOthers {.}}{%
{\protect \APACyear {2010}}%
}]{%
bacon10}
\APACinsertmetastar {%
bacon10}%
\begin{APACrefauthors}%
{Bacon}, R.%
, {Accardo}, M.%
, {Adjali}, L.%
, {Anwand}, H.%
, {Bauer}, S.%
, {Biswas}, I.%
\BDBL {}{Yerle}, N.%
\end{APACrefauthors}%
\unskip\
\newblock
\APACrefYearMonthDay{2010}{{\APACmonth{07}}}{}.
\newblock
{\BBOQ}\APACrefatitle {{The MUSE second-generation VLT instrument}} {{The MUSE
  second-generation VLT instrument}}.{\BBCQ}
\newblock
\BIn{} \APACrefbtitle {\procspie} {\procspie}\ (\BVOL\ 7735, \BPG~773508).
\newblock
\begin{APACrefDOI} \doi{10.1117/12.856027} \end{APACrefDOI}
\PrintBackRefs{\CurrentBib}

\bibitem [\protect \citeauthoryear {%
{Borysow}%
, {Borysow}%
\BCBL {}\ \BBA {} {Fu}%
}{%
{Borysow}%
\ \protect \BOthers {.}}{%
{\protect \APACyear {2000}}%
}]{%
borysow00}
\APACinsertmetastar {%
borysow00}%
\begin{APACrefauthors}%
{Borysow}, A.%
, {Borysow}, J.%
\BCBL {}\ \BBA {} {Fu}, Y.%
\end{APACrefauthors}%
\unskip\
\newblock
\APACrefYearMonthDay{2000}{{\APACmonth{06}}}{}.
\newblock
{\BBOQ}\APACrefatitle {{Semi-empirical Model of Collision-Induced Absorption
  Spectra of H $_{2}$-H $_{2}$ Complexes in the Second Overtone Band of
  Hydrogen at Temperatures from 50 to 500 K}} {{Semi-empirical Model of
  Collision-Induced Absorption Spectra of H $_{2}$-H $_{2}$ Complexes in the
  Second Overtone Band of Hydrogen at Temperatures from 50 to 500 K}}.{\BBCQ}
\newblock
\APACjournalVolNumPages{\icarus}{145}{2}{601-608}.
\newblock
\begin{APACrefDOI} \doi{10.1006/icar.2000.6384} \end{APACrefDOI}
\PrintBackRefs{\CurrentBib}

\bibitem [\protect \citeauthoryear {%
{Borysow}%
\ \BBA {} {Frommhold}%
}{%
{Borysow}%
\ \BBA {} {Frommhold}%
}{%
{\protect \APACyear {1989}}%
}]{%
borysow89b}
\APACinsertmetastar {%
borysow89b}%
\begin{APACrefauthors}%
{Borysow}, A.%
\BCBT {}\ \BBA {} {Frommhold}, L.%
\end{APACrefauthors}%
\unskip\
\newblock
\APACrefYearMonthDay{1989}{{\APACmonth{06}}}{}.
\newblock
{\BBOQ}\APACrefatitle {{Collision-induced Infrared Spectra of H 2-He Pairs at
  Temperatures from 18 to 7000 K. II. Overtone and Hot Bands}}
  {{Collision-induced Infrared Spectra of H 2-He Pairs at Temperatures from 18
  to 7000 K. II. Overtone and Hot Bands}}.{\BBCQ}
\newblock
\APACjournalVolNumPages{\apj}{341}{}{549}.
\newblock
\begin{APACrefDOI} \doi{10.1086/167515} \end{APACrefDOI}
\PrintBackRefs{\CurrentBib}

\bibitem [\protect \citeauthoryear {%
{Borysow}%
, {Frommhold}%
\BCBL {}\ \BBA {} {Moraldi}%
}{%
{Borysow}%
\ \protect \BOthers {.}}{%
{\protect \APACyear {1989}}%
}]{%
borysow89a}
\APACinsertmetastar {%
borysow89a}%
\begin{APACrefauthors}%
{Borysow}, A.%
, {Frommhold}, L.%
\BCBL {}\ \BBA {} {Moraldi}, M.%
\end{APACrefauthors}%
\unskip\
\newblock
\APACrefYearMonthDay{1989}{{\APACmonth{01}}}{}.
\newblock
{\BBOQ}\APACrefatitle {{Collision-induced Infrared Spectra of H 2-He Pairs
  Involving 0 1 Vibrational Transitions and Temperatures from 18 to 7000 K}}
  {{Collision-induced Infrared Spectra of H 2-He Pairs Involving 0 1
  Vibrational Transitions and Temperatures from 18 to 7000 K}}.{\BBCQ}
\newblock
\APACjournalVolNumPages{\apj}{336}{}{495}.
\newblock
\begin{APACrefDOI} \doi{10.1086/167027} \end{APACrefDOI}
\PrintBackRefs{\CurrentBib}

\bibitem [\protect \citeauthoryear {%
{Chance}%
\ \BBA {} {Kurucz}%
}{%
{Chance}%
\ \BBA {} {Kurucz}%
}{%
{\protect \APACyear {2010}}%
}]{%
chance10}
\APACinsertmetastar {%
chance10}%
\begin{APACrefauthors}%
{Chance}, K.%
\BCBT {}\ \BBA {} {Kurucz}, R\BPBI L.%
\end{APACrefauthors}%
\unskip\
\newblock
\APACrefYearMonthDay{2010}{{\APACmonth{06}}}{}.
\newblock
{\BBOQ}\APACrefatitle {{An improved high-resolution solar reference spectrum
  for earth's atmosphere measurements in the ultraviolet, visible, and near
  infrared}} {{An improved high-resolution solar reference spectrum for earth's
  atmosphere measurements in the ultraviolet, visible, and near
  infrared}}.{\BBCQ}
\newblock
\APACjournalVolNumPages{\jqsrt}{111}{9}{1289-1295}.
\newblock
\begin{APACrefDOI} \doi{10.1016/j.jqsrt.2010.01.036} \end{APACrefDOI}
\PrintBackRefs{\CurrentBib}

\bibitem [\protect \citeauthoryear {%
{Grundy}%
, {Schmitt}%
\BCBL {}\ \BBA {} {Quirico}%
}{%
{Grundy}%
\ \protect \BOthers {.}}{%
{\protect \APACyear {2002}}%
}]{%
grundy02}
\APACinsertmetastar {%
grundy02}%
\begin{APACrefauthors}%
{Grundy}, W\BPBI M.%
, {Schmitt}, B.%
\BCBL {}\ \BBA {} {Quirico}, E.%
\end{APACrefauthors}%
\unskip\
\newblock
\APACrefYearMonthDay{2002}{{\APACmonth{02}}}{}.
\newblock
{\BBOQ}\APACrefatitle {{The Temperature-Dependent Spectrum of Methane Ice I
  between 0.7 and 5 {\ensuremath{\mu}}m and Opportunities for Near-Infrared
  Remote Thermometry}} {{The Temperature-Dependent Spectrum of Methane Ice I
  between 0.7 and 5 {\ensuremath{\mu}}m and Opportunities for Near-Infrared
  Remote Thermometry}}.{\BBCQ}
\newblock
\APACjournalVolNumPages{\icarus}{155}{2}{486-496}.
\newblock
\begin{APACrefDOI} \doi{10.1006/icar.2001.6726} \end{APACrefDOI}
\PrintBackRefs{\CurrentBib}

\bibitem [\protect \citeauthoryear {%
{Irwin}%
\ \protect \BOthers {.}}{%
{Irwin}%
\ \protect \BOthers {.}}{%
{\protect \APACyear {2016}}%
}]{%
irwin16}
\APACinsertmetastar {%
irwin16}%
\begin{APACrefauthors}%
{Irwin}, P\BPBI G\BPBI J.%
, {Fletcher}, L\BPBI N.%
, {Tice}, D.%
, {Owen}, S\BPBI J.%
, {Orton}, G\BPBI S.%
, {Teanby}, N\BPBI A.%
\BCBL {}\ \BBA {} {Davis}, G\BPBI R.%
\end{APACrefauthors}%
\unskip\
\newblock
\APACrefYearMonthDay{2016}{{\APACmonth{06}}}{}.
\newblock
{\BBOQ}\APACrefatitle {{Time variability of Neptune's horizontal and vertical
  cloud structure revealed by VLT/SINFONI and Gemini/NIFS from 2009 to 2013}}
  {{Time variability of Neptune's horizontal and vertical cloud structure
  revealed by VLT/SINFONI and Gemini/NIFS from 2009 to 2013}}.{\BBCQ}
\newblock
\APACjournalVolNumPages{\icarus}{271}{}{418-437}.
\newblock
\begin{APACrefDOI} \doi{10.1016/j.icarus.2016.01.015} \end{APACrefDOI}
\PrintBackRefs{\CurrentBib}

\bibitem [\protect \citeauthoryear {%
{Irwin}%
\ \protect \BOthers {.}}{%
{Irwin}%
\ \protect \BOthers {.}}{%
{\protect \APACyear {2011}}%
}]{%
irwin11}
\APACinsertmetastar {%
irwin11}%
\begin{APACrefauthors}%
{Irwin}, P\BPBI G\BPBI J.%
, {Teanby}, N\BPBI A.%
, {Davis}, G\BPBI R.%
, {Fletcher}, L\BPBI N.%
, {Orton}, G\BPBI S.%
, {Tice}, D.%
\BDBL {}{Calcutt}, S\BPBI B.%
\end{APACrefauthors}%
\unskip\
\newblock
\APACrefYearMonthDay{2011}{{\APACmonth{11}}}{}.
\newblock
{\BBOQ}\APACrefatitle {{Multispectral imaging observations of Neptune's cloud
  structure with Gemini-North}} {{Multispectral imaging observations of
  Neptune's cloud structure with Gemini-North}}.{\BBCQ}
\newblock
\APACjournalVolNumPages{\icarus}{216}{1}{141-158}.
\newblock
\begin{APACrefDOI} \doi{10.1016/j.icarus.2011.08.005} \end{APACrefDOI}
\PrintBackRefs{\CurrentBib}

\bibitem [\protect \citeauthoryear {%
{Irwin}%
\ \protect \BOthers {.}}{%
{Irwin}%
\ \protect \BOthers {.}}{%
{\protect \APACyear {2008}}%
}]{%
irwin08}
\APACinsertmetastar {%
irwin08}%
\begin{APACrefauthors}%
{Irwin}, P\BPBI G\BPBI J.%
, {Teanby}, N\BPBI A.%
, {de Kok}, R.%
, {Fletcher}, L\BPBI N.%
, {Howett}, C\BPBI J\BPBI A.%
, {Tsang}, C\BPBI C\BPBI C.%
\BDBL {}{Parrish}, P\BPBI D.%
\end{APACrefauthors}%
\unskip\
\newblock
\APACrefYearMonthDay{2008}{{\APACmonth{04}}}{}.
\newblock
{\BBOQ}\APACrefatitle {{The NEMESIS planetary atmosphere radiative transfer and
  retrieval tool}} {{The NEMESIS planetary atmosphere radiative transfer and
  retrieval tool}}.{\BBCQ}
\newblock
\APACjournalVolNumPages{\jqsrt}{109}{}{1136-1150}.
\newblock
\begin{APACrefDOI} \doi{10.1016/j.jqsrt.2007.11.006} \end{APACrefDOI}
\PrintBackRefs{\CurrentBib}

\bibitem [\protect \citeauthoryear {%
{Irwin}%
\ \protect \BOthers {.}}{%
{Irwin}%
\ \protect \BOthers {.}}{%
{\protect \APACyear {2019}}%
}]{%
irwin19}
\APACinsertmetastar {%
irwin19}%
\begin{APACrefauthors}%
{Irwin}, P\BPBI G\BPBI J.%
, {Toledo}, D.%
, {Braude}, A\BPBI S.%
, {Bacon}, R.%
, {Weilbacher}, P\BPBI M.%
, {Teanby}, N\BPBI A.%
\BDBL {}{Orton}, G\BPBI S.%
\end{APACrefauthors}%
\unskip\
\newblock
\APACrefYearMonthDay{2019}{{\APACmonth{10}}}{}.
\newblock
{\BBOQ}\APACrefatitle {{Latitudinal variation in the abundance of methane
  (CH$_{4}$) above the clouds in Neptune's atmosphere from VLT/MUSE Narrow
  Field Mode Observations}} {{Latitudinal variation in the abundance of methane
  (CH$_{4}$) above the clouds in Neptune's atmosphere from VLT/MUSE Narrow
  Field Mode Observations}}.{\BBCQ}
\newblock
\APACjournalVolNumPages{\icarus}{331}{}{69-82}.
\newblock
\begin{APACrefDOI} \doi{10.1016/j.icarus.2019.05.011} \end{APACrefDOI}
\PrintBackRefs{\CurrentBib}

\bibitem [\protect \citeauthoryear {%
{Irwin}%
\ \protect \BOthers {.}}{%
{Irwin}%
\ \protect \BOthers {.}}{%
{\protect \APACyear {2018}}%
}]{%
irwin18}
\APACinsertmetastar {%
irwin18}%
\begin{APACrefauthors}%
{Irwin}, P\BPBI G\BPBI J.%
, {Toledo}, D.%
, {Garland}, R.%
, {Teanby}, N\BPBI A.%
, {Fletcher}, L\BPBI N.%
, {Orton}, G\BPBI S.%
\BCBL {}\ \BBA {} {B{\'e}zard}, B.%
\end{APACrefauthors}%
\unskip\
\newblock
\APACrefYearMonthDay{2018}{{\APACmonth{04}}}{}.
\newblock
{\BBOQ}\APACrefatitle {{Detection of hydrogen sulfide above the clouds in
  Uranus's atmosphere}} {{Detection of hydrogen sulfide above the clouds in
  Uranus's atmosphere}}.{\BBCQ}
\newblock
\APACjournalVolNumPages{Nature Astronomy}{2}{}{420-427}.
\newblock
\begin{APACrefDOI} \doi{10.1038/s41550-018-0432-1} \end{APACrefDOI}
\PrintBackRefs{\CurrentBib}

\bibitem [\protect \citeauthoryear {%
{Karkoschka}%
\ \BBA {} {Tomasko}%
}{%
{Karkoschka}%
\ \BBA {} {Tomasko}%
}{%
{\protect \APACyear {2009}}%
}]{%
kark09}
\APACinsertmetastar {%
kark09}%
\begin{APACrefauthors}%
{Karkoschka}, E.%
\BCBT {}\ \BBA {} {Tomasko}, M.%
\end{APACrefauthors}%
\unskip\
\newblock
\APACrefYearMonthDay{2009}{{\APACmonth{07}}}{}.
\newblock
{\BBOQ}\APACrefatitle {{The haze and methane distributions on Uranus from
  HST-STIS spectroscopy}} {{The haze and methane distributions on Uranus from
  HST-STIS spectroscopy}}.{\BBCQ}
\newblock
\APACjournalVolNumPages{\icarus}{202}{1}{287-309}.
\newblock
\begin{APACrefDOI} \doi{10.1016/j.icarus.2009.02.010} \end{APACrefDOI}
\PrintBackRefs{\CurrentBib}

\bibitem [\protect \citeauthoryear {%
{Karkoschka}%
\ \BBA {} {Tomasko}%
}{%
{Karkoschka}%
\ \BBA {} {Tomasko}%
}{%
{\protect \APACyear {2010}}%
}]{%
kark10}
\APACinsertmetastar {%
kark10}%
\begin{APACrefauthors}%
{Karkoschka}, E.%
\BCBT {}\ \BBA {} {Tomasko}, M\BPBI G.%
\end{APACrefauthors}%
\unskip\
\newblock
\APACrefYearMonthDay{2010}{{\APACmonth{02}}}{}.
\newblock
{\BBOQ}\APACrefatitle {{Methane absorption coefficients for the jovian planets
  from laboratory, Huygens, and HST data}} {{Methane absorption coefficients
  for the jovian planets from laboratory, Huygens, and HST data}}.{\BBCQ}
\newblock
\APACjournalVolNumPages{\icarus}{205}{2}{674-694}.
\newblock
\begin{APACrefDOI} \doi{10.1016/j.icarus.2009.07.044} \end{APACrefDOI}
\PrintBackRefs{\CurrentBib}

\bibitem [\protect \citeauthoryear {%
{Karkoschka}%
\ \BBA {} {Tomasko}%
}{%
{Karkoschka}%
\ \BBA {} {Tomasko}%
}{%
{\protect \APACyear {2011}}%
}]{%
kark11}
\APACinsertmetastar {%
kark11}%
\begin{APACrefauthors}%
{Karkoschka}, E.%
\BCBT {}\ \BBA {} {Tomasko}, M\BPBI G.%
\end{APACrefauthors}%
\unskip\
\newblock
\APACrefYearMonthDay{2011}{{\APACmonth{01}}}{}.
\newblock
{\BBOQ}\APACrefatitle {{The haze and methane distributions on Neptune from
  HST-STIS spectroscopy}} {{The haze and methane distributions on Neptune from
  HST-STIS spectroscopy}}.{\BBCQ}
\newblock
\APACjournalVolNumPages{\icarus}{211}{1}{780-797}.
\newblock
\begin{APACrefDOI} \doi{10.1016/j.icarus.2010.08.013} \end{APACrefDOI}
\PrintBackRefs{\CurrentBib}

\bibitem [\protect \citeauthoryear {%
{Lellouch}%
\ \protect \BOthers {.}}{%
{Lellouch}%
\ \protect \BOthers {.}}{%
{\protect \APACyear {2010}}%
}]{%
lellouch10}
\APACinsertmetastar {%
lellouch10}%
\begin{APACrefauthors}%
{Lellouch}, E.%
, {Hartogh}, P.%
, {Feuchtgruber}, H.%
, {Vand enbussche}, B.%
, {de Graauw}, T.%
, {Moreno}, R.%
\BDBL {}{Wildeman}, K.%
\end{APACrefauthors}%
\unskip\
\newblock
\APACrefYearMonthDay{2010}{{\APACmonth{07}}}{}.
\newblock
{\BBOQ}\APACrefatitle {{First results of Herschel-PACS observations of
  Neptune}} {{First results of Herschel-PACS observations of Neptune}}.{\BBCQ}
\newblock
\APACjournalVolNumPages{\aap}{518}{}{L152}.
\newblock
\begin{APACrefDOI} \doi{10.1051/0004-6361/201014600} \end{APACrefDOI}
\PrintBackRefs{\CurrentBib}

\bibitem [\protect \citeauthoryear {%
{Lindal}%
}{%
{Lindal}%
}{%
{\protect \APACyear {1992}}%
}]{%
lindal92}
\APACinsertmetastar {%
lindal92}%
\begin{APACrefauthors}%
{Lindal}, G\BPBI F.%
\end{APACrefauthors}%
\unskip\
\newblock
\APACrefYearMonthDay{1992}{{\APACmonth{03}}}{}.
\newblock
{\BBOQ}\APACrefatitle {{The Atmosphere of Neptune: an Analysis of Radio
  Occultation Data Acquired with Voyager 2}} {{The Atmosphere of Neptune: an
  Analysis of Radio Occultation Data Acquired with Voyager 2}}.{\BBCQ}
\newblock
\APACjournalVolNumPages{\aj}{103}{}{967}.
\newblock
\begin{APACrefDOI} \doi{10.1086/116119} \end{APACrefDOI}
\PrintBackRefs{\CurrentBib}

\bibitem [\protect \citeauthoryear {%
{Luszcz-Cook}%
, {de Kleer}%
, {de Pater}%
, {Adamkovics}%
\BCBL {}\ \BBA {} {Hammel}%
}{%
{Luszcz-Cook}%
\ \protect \BOthers {.}}{%
{\protect \APACyear {2016}}%
}]{%
luszcz16}
\APACinsertmetastar {%
luszcz16}%
\begin{APACrefauthors}%
{Luszcz-Cook}, S\BPBI H.%
, {de Kleer}, K.%
, {de Pater}, I.%
, {Adamkovics}, M.%
\BCBL {}\ \BBA {} {Hammel}, H\BPBI B.%
\end{APACrefauthors}%
\unskip\
\newblock
\APACrefYearMonthDay{2016}{{\APACmonth{09}}}{}.
\newblock
{\BBOQ}\APACrefatitle {{Retrieving Neptune's aerosol properties from Keck
  OSIRIS observations. I. Dark regions}} {{Retrieving Neptune's aerosol
  properties from Keck OSIRIS observations. I. Dark regions}}.{\BBCQ}
\newblock
\APACjournalVolNumPages{\icarus}{276}{}{52-87}.
\newblock
\begin{APACrefDOI} \doi{10.1016/j.icarus.2016.04.032} \end{APACrefDOI}
\PrintBackRefs{\CurrentBib}

\bibitem [\protect \citeauthoryear {%
{Martonchik}%
\ \BBA {} {Orton}%
}{%
{Martonchik}%
\ \BBA {} {Orton}%
}{%
{\protect \APACyear {1994}}%
}]{%
martonchick94}
\APACinsertmetastar {%
martonchick94}%
\begin{APACrefauthors}%
{Martonchik}, J\BPBI V.%
\BCBT {}\ \BBA {} {Orton}, G\BPBI S.%
\end{APACrefauthors}%
\unskip\
\newblock
\APACrefYearMonthDay{1994}{{\APACmonth{12}}}{}.
\newblock
{\BBOQ}\APACrefatitle {{Optical constants of liquid and solid methane}}
  {{Optical constants of liquid and solid methane}}.{\BBCQ}
\newblock
\APACjournalVolNumPages{\ao}{33}{36}{8306-8317}.
\newblock
\begin{APACrefDOI} \doi{10.1364/AO.33.008306} \end{APACrefDOI}
\PrintBackRefs{\CurrentBib}

\bibitem [\protect \citeauthoryear {%
{Minnaert}%
}{%
{Minnaert}%
}{%
{\protect \APACyear {1941}}%
}]{%
minnaert41}
\APACinsertmetastar {%
minnaert41}%
\begin{APACrefauthors}%
{Minnaert}, M.%
\end{APACrefauthors}%
\unskip\
\newblock
\APACrefYearMonthDay{1941}{}{}.
\newblock
{\BBOQ}\APACrefatitle {{The reciprocity principle in lunar photometry}} {{The
  reciprocity principle in lunar photometry}}.{\BBCQ}
\newblock
\APACjournalVolNumPages{\apj}{93}{}{403-410}.
\newblock
\begin{APACrefDOI} \doi{10.1086/144279} \end{APACrefDOI}
\PrintBackRefs{\CurrentBib}

\bibitem [\protect \citeauthoryear {%
{P{\'e}rez-Hoyos}%
\ \protect \BOthers {.}}{%
{P{\'e}rez-Hoyos}%
\ \protect \BOthers {.}}{%
{\protect \APACyear {2020}}%
}]{%
perez20}
\APACinsertmetastar {%
perez20}%
\begin{APACrefauthors}%
{P{\'e}rez-Hoyos}, S.%
, {S{\'a}nchez-Lavega}, A.%
, {Sanz-Requena}, J\BPBI F.%
, {Barrado-Izagirre}, N.%
, {Carri{\'o}n-Gonz{\'a}lez}, O.%
, {Anguiano-Arteaga}, A.%
\BDBL {}{Braude}, A\BPBI S.%
\end{APACrefauthors}%
\unskip\
\newblock
\APACrefYearMonthDay{2020}{{\APACmonth{12}}}{}.
\newblock
{\BBOQ}\APACrefatitle {{Color and aerosol changes in Jupiter after a North
  Temperate Belt disturbance}} {{Color and aerosol changes in Jupiter after a
  North Temperate Belt disturbance}}.{\BBCQ}
\newblock
\APACjournalVolNumPages{\icarus}{352}{}{114031}.
\newblock
\begin{APACrefDOI} \doi{10.1016/j.icarus.2020.114031} \end{APACrefDOI}
\PrintBackRefs{\CurrentBib}

\bibitem [\protect \citeauthoryear {%
{Plass}%
, {Kattawar}%
\BCBL {}\ \BBA {} {Catchings}%
}{%
{Plass}%
\ \protect \BOthers {.}}{%
{\protect \APACyear {1973}}%
}]{%
plass73}
\APACinsertmetastar {%
plass73}%
\begin{APACrefauthors}%
{Plass}, G\BPBI N.%
, {Kattawar}, G\BPBI W.%
\BCBL {}\ \BBA {} {Catchings}, F\BPBI E.%
\end{APACrefauthors}%
\unskip\
\newblock
\APACrefYearMonthDay{1973}{{\APACmonth{01}}}{}.
\newblock
{\BBOQ}\APACrefatitle {{Matrix operator theory of radiative transfer. 1:
  Rayleigh scattering.}} {{Matrix operator theory of radiative transfer. 1:
  Rayleigh scattering.}}{\BBCQ}
\newblock
\APACjournalVolNumPages{\ao}{12}{}{314-329}.
\newblock
\begin{APACrefDOI} \doi{10.1364/AO.12.000314} \end{APACrefDOI}
\PrintBackRefs{\CurrentBib}

\bibitem [\protect \citeauthoryear {%
{Sromovsky}%
, {Fry}%
\BCBL {}\ \BBA {} {Kim}%
}{%
{Sromovsky}%
\ \protect \BOthers {.}}{%
{\protect \APACyear {2011}}%
}]{%
sromovsky11}
\APACinsertmetastar {%
sromovsky11}%
\begin{APACrefauthors}%
{Sromovsky}, L\BPBI A.%
, {Fry}, P\BPBI M.%
\BCBL {}\ \BBA {} {Kim}, J\BPBI H.%
\end{APACrefauthors}%
\unskip\
\newblock
\APACrefYearMonthDay{2011}{{\APACmonth{09}}}{}.
\newblock
{\BBOQ}\APACrefatitle {{Methane on Uranus: The case for a compact CH$_{4}$
  cloud layer at low latitudes and a severe CH$_{4}$ depletion at
  high-latitudes based on re-analysis of Voyager occultation measurements and
  STIS spectroscopy}} {{Methane on Uranus: The case for a compact CH$_{4}$
  cloud layer at low latitudes and a severe CH$_{4}$ depletion at
  high-latitudes based on re-analysis of Voyager occultation measurements and
  STIS spectroscopy}}.{\BBCQ}
\newblock
\APACjournalVolNumPages{\icarus}{215}{1}{292-312}.
\newblock
\begin{APACrefDOI} \doi{10.1016/j.icarus.2011.06.024} \end{APACrefDOI}
\PrintBackRefs{\CurrentBib}

\bibitem [\protect \citeauthoryear {%
{Sromovsky}%
, {Karkoschka}%
, {Fry}%
, {de Pater}%
\BCBL {}\ \BBA {} {Hammel}%
}{%
{Sromovsky}%
\ \protect \BOthers {.}}{%
{\protect \APACyear {2019}}%
}]{%
sromovsky19}
\APACinsertmetastar {%
sromovsky19}%
\begin{APACrefauthors}%
{Sromovsky}, L\BPBI A.%
, {Karkoschka}, E.%
, {Fry}, P\BPBI M.%
, {de Pater}, I.%
\BCBL {}\ \BBA {} {Hammel}, H\BPBI B.%
\end{APACrefauthors}%
\unskip\
\newblock
\APACrefYearMonthDay{2019}{{\APACmonth{01}}}{}.
\newblock
{\BBOQ}\APACrefatitle {{The methane distribution and polar brightening on
  Uranus based on HST/STIS, Keck/NIRC2, and IRTF/SpeX observations through
  2015}} {{The methane distribution and polar brightening on Uranus based on
  HST/STIS, Keck/NIRC2, and IRTF/SpeX observations through 2015}}.{\BBCQ}
\newblock
\APACjournalVolNumPages{\icarus}{317}{}{266-306}.
\newblock
\begin{APACrefDOI} \doi{10.1016/j.icarus.2018.06.026} \end{APACrefDOI}
\PrintBackRefs{\CurrentBib}

\bibitem [\protect \citeauthoryear {%
{Sromovsky}%
\ \protect \BOthers {.}}{%
{Sromovsky}%
\ \protect \BOthers {.}}{%
{\protect \APACyear {2014}}%
}]{%
sromovsky14}
\APACinsertmetastar {%
sromovsky14}%
\begin{APACrefauthors}%
{Sromovsky}, L\BPBI A.%
, {Karkoschka}, E.%
, {Fry}, P\BPBI M.%
, {Hammel}, H\BPBI B.%
, {de Pater}, I.%
\BCBL {}\ \BBA {} {Rages}, K.%
\end{APACrefauthors}%
\unskip\
\newblock
\APACrefYearMonthDay{2014}{{\APACmonth{08}}}{}.
\newblock
{\BBOQ}\APACrefatitle {{Methane depletion in both polar regions of Uranus
  inferred from HST/STIS and Keck/NIRC2 observations}} {{Methane depletion in
  both polar regions of Uranus inferred from HST/STIS and Keck/NIRC2
  observations}}.{\BBCQ}
\newblock
\APACjournalVolNumPages{\icarus}{238}{}{137-155}.
\newblock
\begin{APACrefDOI} \doi{10.1016/j.icarus.2014.05.016} \end{APACrefDOI}
\PrintBackRefs{\CurrentBib}

\bibitem [\protect \citeauthoryear {%
{Tollefson}%
, {de Pater}%
, {Luszcz-Cook}%
\BCBL {}\ \BBA {} {DeBoer}%
}{%
{Tollefson}%
\ \protect \BOthers {.}}{%
{\protect \APACyear {2019}}%
}]{%
tollefson19}
\APACinsertmetastar {%
tollefson19}%
\begin{APACrefauthors}%
{Tollefson}, J.%
, {de Pater}, I.%
, {Luszcz-Cook}, S.%
\BCBL {}\ \BBA {} {DeBoer}, D.%
\end{APACrefauthors}%
\unskip\
\newblock
\APACrefYearMonthDay{2019}{{\APACmonth{06}}}{}.
\newblock
{\BBOQ}\APACrefatitle {{Neptune's Latitudinal Variations as Viewed with ALMA}}
  {{Neptune's Latitudinal Variations as Viewed with ALMA}}.{\BBCQ}
\newblock
\APACjournalVolNumPages{\aj}{157}{6}{251}.
\newblock
\begin{APACrefDOI} \doi{10.3847/1538-3881/ab1fdf} \end{APACrefDOI}
\PrintBackRefs{\CurrentBib}

\end{thebibliography}

%
%
%
%
%

\end{document}